\documentclass{mn2e}

\usepackage{epsfig}
\usepackage{rotating}
\usepackage{graphicx}
\usepackage{epstopdf}
\newif\ifAMStwofonts
\AMStwofontstrue

\title[The First Source Counts at 18 microns]{The First Source Counts at 18 microns from the AKARI NEP Survey }
\author[C.~Pearson  {\it et al.}]
       {
       Chris P. Pearson$^1$$^{,2,8}$\thanks{For further information please contact Chris Pearson (chris.pearson@stfc.ac.uk)},
S.~Serjeant$^2$,
S.~Oyabu$^7$,
H.~Matsuhara$^3$, 
T.~Wada$^3$,
\vspace*{0.3cm}\\ 
{\LARGE  \textup{
T.~Goto$^5$,
T.~Takagi$^3$, 
H.M.~Lee$^4$, 
M.~Im$^4$, 
Y.~Ohyama$^6$, 
S.J.~Kim$^4$,
K.~Murata$^3$
}}
\vspace*{0.1cm}\\
      $^1$ RAL Space, CCLRC Rutherford Appleton Laboratory, Chilton, Didcot, Oxfordshire OX11 0QX, United Kingdom\\
      $^2$ Department of Physical Sciences, The Open University, Milton Keynes, MK7 6AA, UK\\
      $^3$ Institute of Space and Astronautical Science, Yoshinodai 3-1-1, Sagamihara, Kanagawa 229 8510, Japan\\
      $^4$ Department of Physics and Astronomy, Seoul National University, Shillim-Dong Kwanak-Gu, Seoul 151-742, South Korea \\
      $^5$ Dark Cosmology Centre, Niels Bohr Institute, University of Copenhagen, Denmark \\
      $^6$ Institute of Astronomy and Astrophysics Academia Sinica, Taipei 10617, Taiwan, R.O.C.\\
      $^7$ Graduate School of Science, Nagoya University, Nagoya 464-8602, Japan\\
      $^8$ Oxford Astrophysics, Denys Wilkinson Building, University of Oxford, Keble Rd, Oxford OX1 3RH, UK
      }
\date{Accepted .\\
      Received ;\\
      in original form 2013 May }
\pagerange{\pageref{firstpage}--\pageref{lastpage}}

\begin{document}

\label{firstpage}

\maketitle

\begin{abstract}
We present the first galaxy counts at 18 microns using the Japanese AKARI satellite's survey at the North Ecliptic Pole (NEP), produced from the images from the NEP-Deep and NEP-Wide surveys covering 0.6 and 5.8 square degrees respectively. We describe a procedure  using a point source filtering algorithm to remove background structure and a minimum variance method for our source extraction and photometry that delivers the optimum signal to noise for our extracted sources, confirming this by comparison with  standard photometry methods.  The final source counts are complete and reliable over three orders of magnitude in flux density, resulting in sensitivities (80 percent completeness) of 0.15mJy and 0.3mJy for the  NEP-Deep and NEP-Wide surveys respectively, a factor of 1.3 deeper than previous catalogues constructed from this field. The differential source counts exhibit a characteristic upturn from Euclidean expectations at around a milliJansky and a corresponding evolutionary bump between 0.2-0.4 mJy consistent with previous mid-infrared surveys with ISO and Spitzer at 15 and 24 microns. We compare our results with galaxy evolution models confirming the  striking  divergence from the non-evolving scenario. The models and observations are in broad agreement implying that the source counts are consistent with a strongly evolving population of luminous infrared galaxies at redshifts higher than unity. Integrating our source counts down to the limit of the NEP survey at the 150 microJy level we calculate that AKARI has resolved approximately 55 percent of the 18 micron cosmic infrared background relative to the predictions of contemporary source count models.  
\end{abstract}

\begin{keywords}
 Infrared: source counts, Surveys -- Cosmology: source counts -- Galaxies: evolution.
\end{keywords}


\section{Introduction}\label{sec:introduction}
 
 At least half the radiative energy output from star formation
 throughout the history of the Universe has been intercepted and
 absorbed by dust, which re-radiated the energy in the infrared 
 (Hauser
 et al. ~\shortcite{hauser98},  Lagache \&
 Puget~\shortcite{lagache00}, Franceschini et
 al. \shortcite{franceschini01}). 
 This 
 emphasizes the requirement for large
 galaxy surveys and galaxy source counts at infrared to submillimetre
 wavelengths to determine the statistical properties of the dominant
 contributing sources.  
 The peak in the dusty star formation
 emission  lies in the wavelength region between 50-200\,$\mu$m, 
 however the observational constraints and difficulties 
 in the far-infrared to submillimetre
 due to detector array size, instrumental effects, resolution, etc,
together with 
the fact that a significant fraction ($\sim$40$\%$) of the
 luminosity in star forming galaxies is radiated in the mid-infrared
 region from 5-50\,$\mu$m, has led to a string of 
 successes for 
 mid-infrared 
 space observatories. 
 
 The many surveys carried out at 15\,$\mu$m with the Infrared Space
 Observatory ({\it ISO}, e.g. Elbaz et al. \shortcite{elbaz99},
 \shortcite{elbaz02}) revealed a rapidly evolving population of
 star-forming galaxies from redshifts $\sim$0.4 -- 1
 (e.g. Serjeant et al. \shortcite{serjeant00}). These sources
 required strong evolution that was unexpected from either optical
 surveys or previous surveys in the infrared, requiring a reassessment
 of galaxy evolution models (e.g. Pearson \shortcite{cpp01}). With the
 launch of {\it Spitzer} in 2003, and the superb performance of the
 MIPS 24\,$\mu$m band, the higher redshift (z=1-3) Universe became
 accessible, confirming the  strong evolution of the mid-infrared
 population (Papovich et al. ~\shortcite{papovich04}, Marleau et
 al. ~\shortcite{marleau04}, Chary et al. ~\shortcite{chary04}). 
 The first surveys at 24\,$\mu$m provided stronger constraints
on models that tracked the evolving galaxy population at high redshifts. 
 These surveys led to a revision in galaxy evolution models which had
 failed to predict such a strongly evolving population at high
 redshifts (e.g. Xu et al. ~\shortcite{xu03}, Chary \& Elbaz
 ~\shortcite{chary03}, King \& Rowan-Robinson ~\shortcite{king03},
 Lagache et al.  ~\shortcite{lagache03}). 
 Pearson \shortcite{cpp05} produced models that simultaneously fit
 both the counts at 15 and 24\,$\mu$m commenting that the assumed
 galaxy spectra and evolution mode were important factors in modelling
 the mid-infrared source counts. The mid-infrared spectra of galaxies
 is dominated by the emission and absorption from dust features at
 3.3, 6.2, 7.7, 8.6, 9.7, 11.3 and 12.7\,$\mu$m which 
 complicate the evolutionary modelling \cite{takagi12a} and
 small changes in the assumed mid-infrared spectra 
can result in large effects on the resulting counts \cite{lagache04}. Given the sensitive nature of the
 mid-infrared, it is extremely important to provide surveys in
 additional bands in order to constrain the evolution of the dusty
 galaxy population to high redshift and to unravel the spectral and
 evolutionary contributions. 
 
The {\it AKARI} satellite  \cite{murakami07} was a Japanese space
mission dedicated to infrared astrophysics and was launched on board
JAXA's M-V8 Launch Vehicle on February 22, 2006 (Japan Standard Time).
 {\it AKARI} had a 68.5\,cm cooled telescope with two focal plane
instruments, the Far-Infrared Surveyor (FIS) \cite{kawada07} and the
Infrared Camera (IRC).  The IRC consists of three cameras, the
IRC-NIR, MIR-S and MIR-L covering 1.7-- 26\,$\mu$m in 9 mid-infrared
bands (seven narrow bands at 2.4, 3.2, 4.1, 7, 11, 15, 24\,$\mu$m and
two additional wide bands at 9 and 18\,$\mu$m ) with fields of view
(FoV) of 10$\arcmin$  $\times$ 10$\arcmin$ \cite{onaka07}. Although
the NIR/MIR-S cameras share the same field of view, the MIR-L camera
is separated on the sky by 20 arcmins. {\it AKARI} observations are
segregated into Large Surveys (LS) and guaranteed time Mission
Programs. Due to the Sun-synchronous nature of {\it AKARI}'s orbit,
deep or wide surveys are naturally constrained to high ecliptic
latitudes and optimally at the opposing North and South Ecliptic
Poles. 
The largest guaranteed  time LS programs have been
carried out at the South Ecliptic Pole (the {\it AKARI} survey of the
Large Magellanic Cloud Ita et al. \shortcite{ita08}, \shortcite{ita12}, and North Ecliptic Pole (The
{\it AKARI} NEP extragalactic survey, Matsuhara et
al. \shortcite{matsuhara06}). 

The NEP survey 
consists of a two tiered survey of overlapping but not concentric circular 
areas: a deep central region covering $\sim$ 0.6 square degrees, centred on
RA=17h56m, Dec.=66d37m, hereafter referred to as the  NEP-Deep survey
\cite{wada08} and a shallow surrounding region covering $\sim$ 5.8
square degrees  centred on the NEP, hereafter referred to as
the NEP-Wide survey \cite{lee09}.  The NEP survey has also been targeted by ESA's Herschel Space Observatory \cite{pilbratt10} for a survey at far-infrared to sub millimetre wavelengths (Serjeant et al. in preparation, Pearson et al. in preparation) and is a  candidate for the {\it Euclid} deep field \cite{serjeant12}

In this paper we report on the construction of the  {\it AKARI}-IRC
$18\,\mu$m band (the L18W band covering the wavelength range 13.9 --
25.6\,$\mu$m) photometry and  source counts   for both the NEP-Deep
and NEP-Wide surveys. The width of the L18W band (encompassing both
the {\it ISO} 15\,$\mu$m and {\it Spitzer} 24\,$\mu$m bands) makes these
surveys less sensitive to the passage of individual dust features through the
pass band. In this work we concentrate on advanced image
processing, source extraction and photometry for the NEP-Deep survey
in order to obtain a robust catalogue with optimal flux densities for sources
down to faint levels. In Section \ref{sec:nepsurvey} we summarize the
observations and data reduction. In Section \ref{sec:processing} we
discuss our re-processing of the original images, and our source extraction method. We discuss the
photometry of our extract sources in Section \ref{sec:photometry} and
compare this with standard aperture photometry techniques. Finally, we
construct the source counts at 18\,$\mu$m in the   {\it AKARI} L18W band
for both the NEP-Deep and NEP-Wide surveys in Section \ref{sec:Source
  Counts}. The conclusions are given in Section
\ref{sec:conclusions}. 
  Throughout this work we assume a Hubble constant of 
  $H_0=72$\,km\,s$^{-1}$\,Mpc$^{-1}$
  and density parameters of $\Omega_{\rm M}=0.3$ and 
  $\Omega_\Lambda=0.7$ from matter and the Cosmological Constant
  respectively, so as to be consistent in comparing with previous studies in the same field.

\section{The NEP survey at 18 microns}\label{sec:nepsurvey}

\subsection{Observations}\label{sec:observations}

\begin{figure*}
\centering
\centerline{
\psfig{ figure=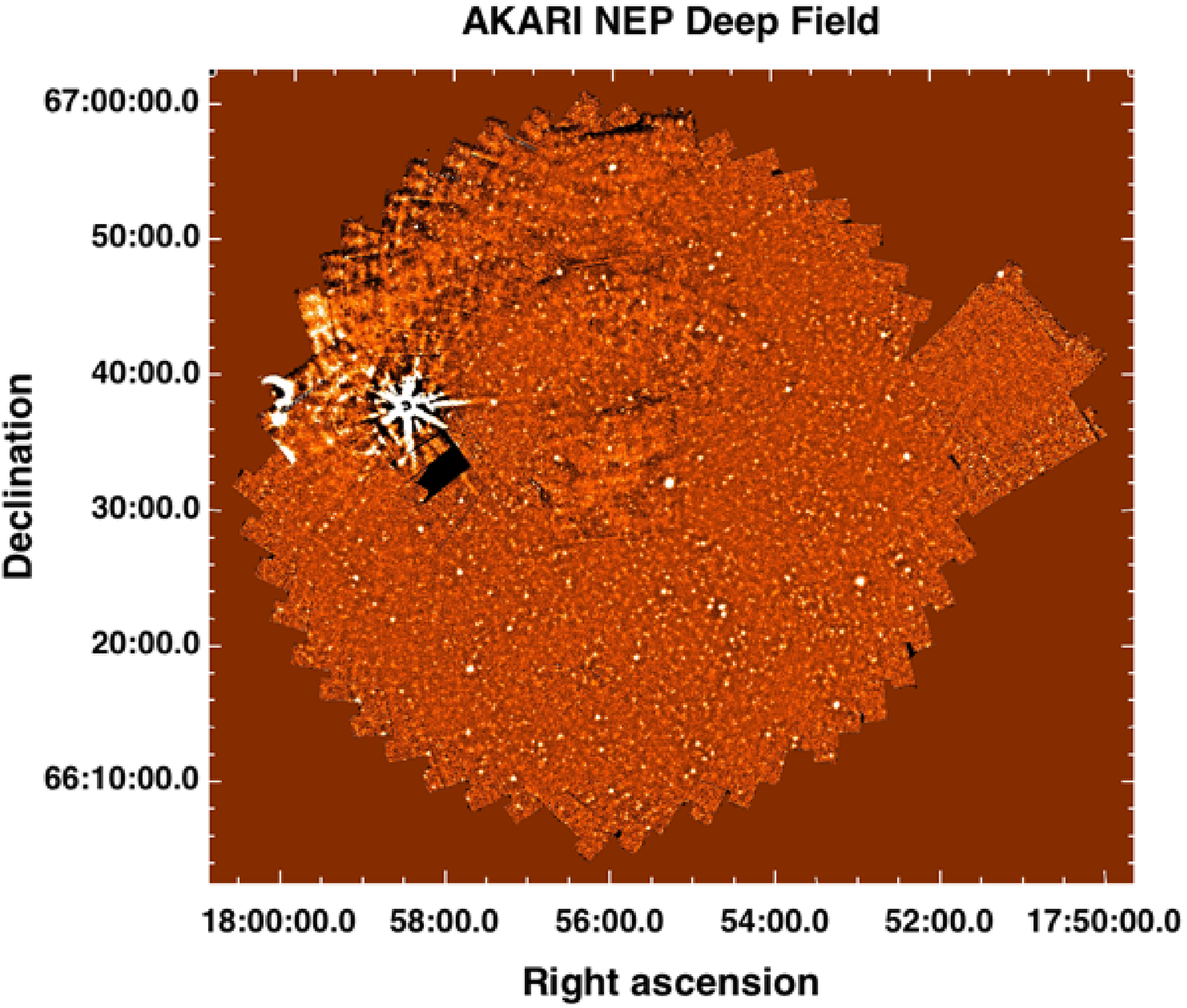,height=8cm}
\psfig{ figure=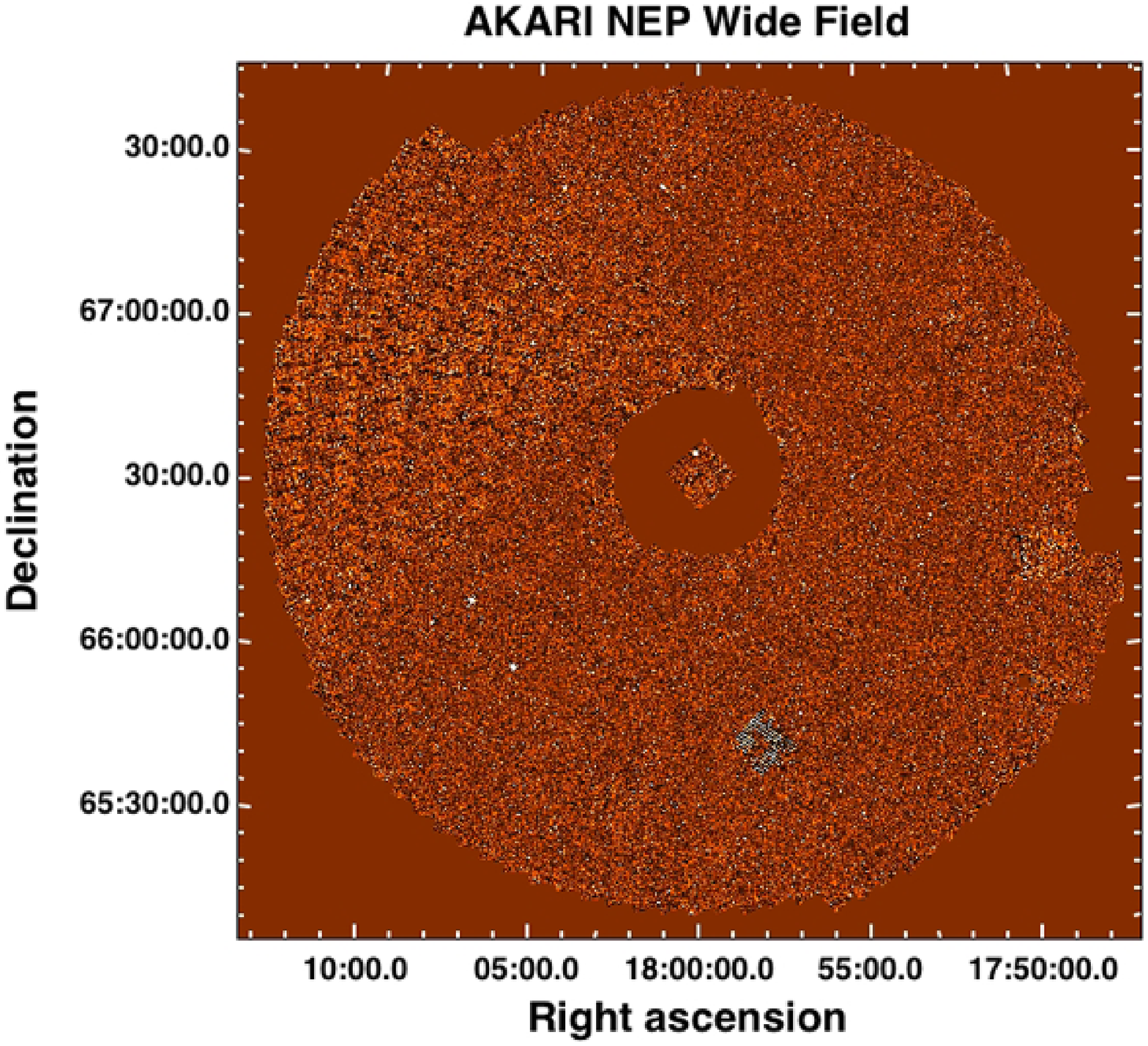,height=8cm}
}
\centerline{
\psfig{ figure=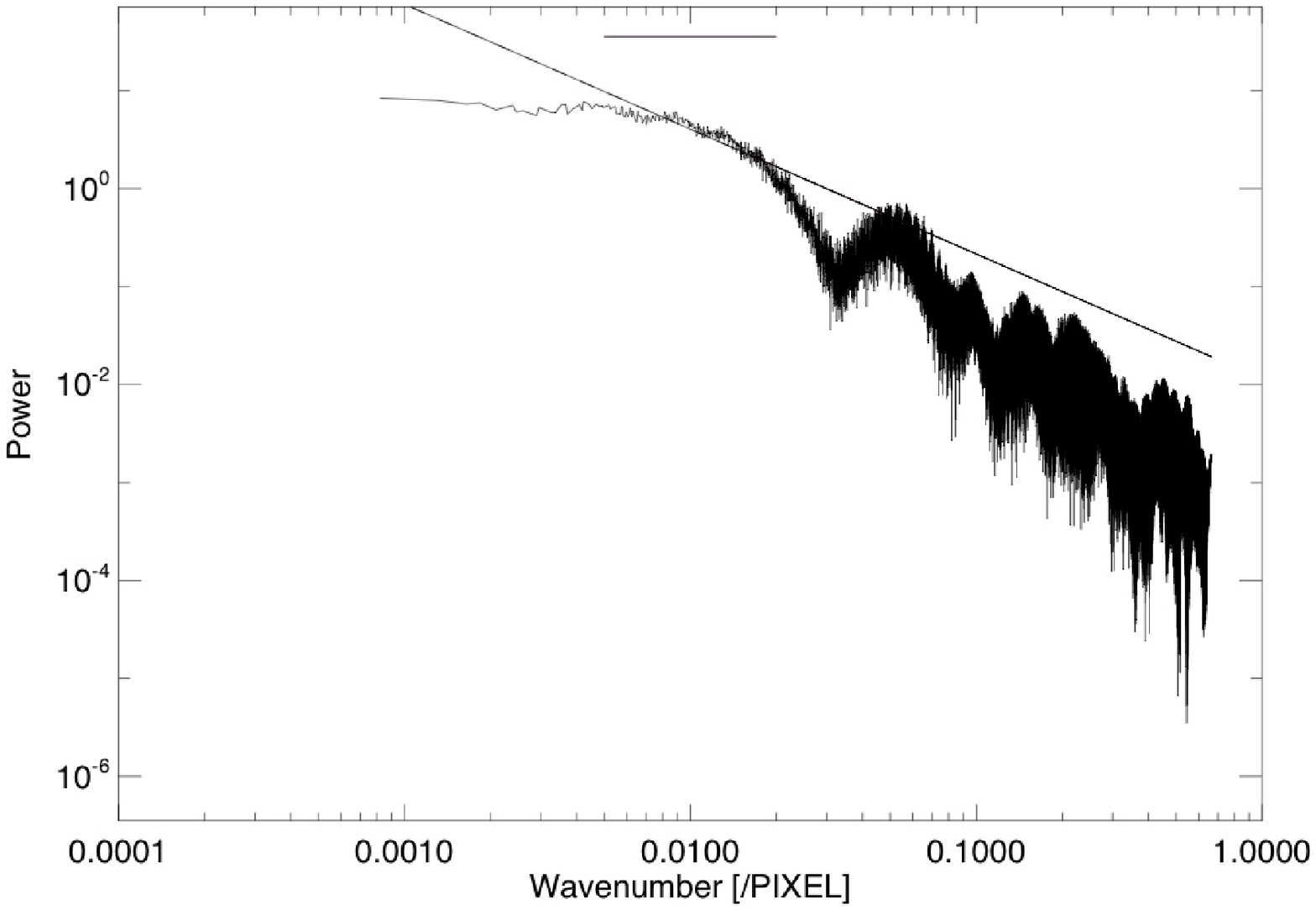,width=8.8cm}
\psfig{ figure=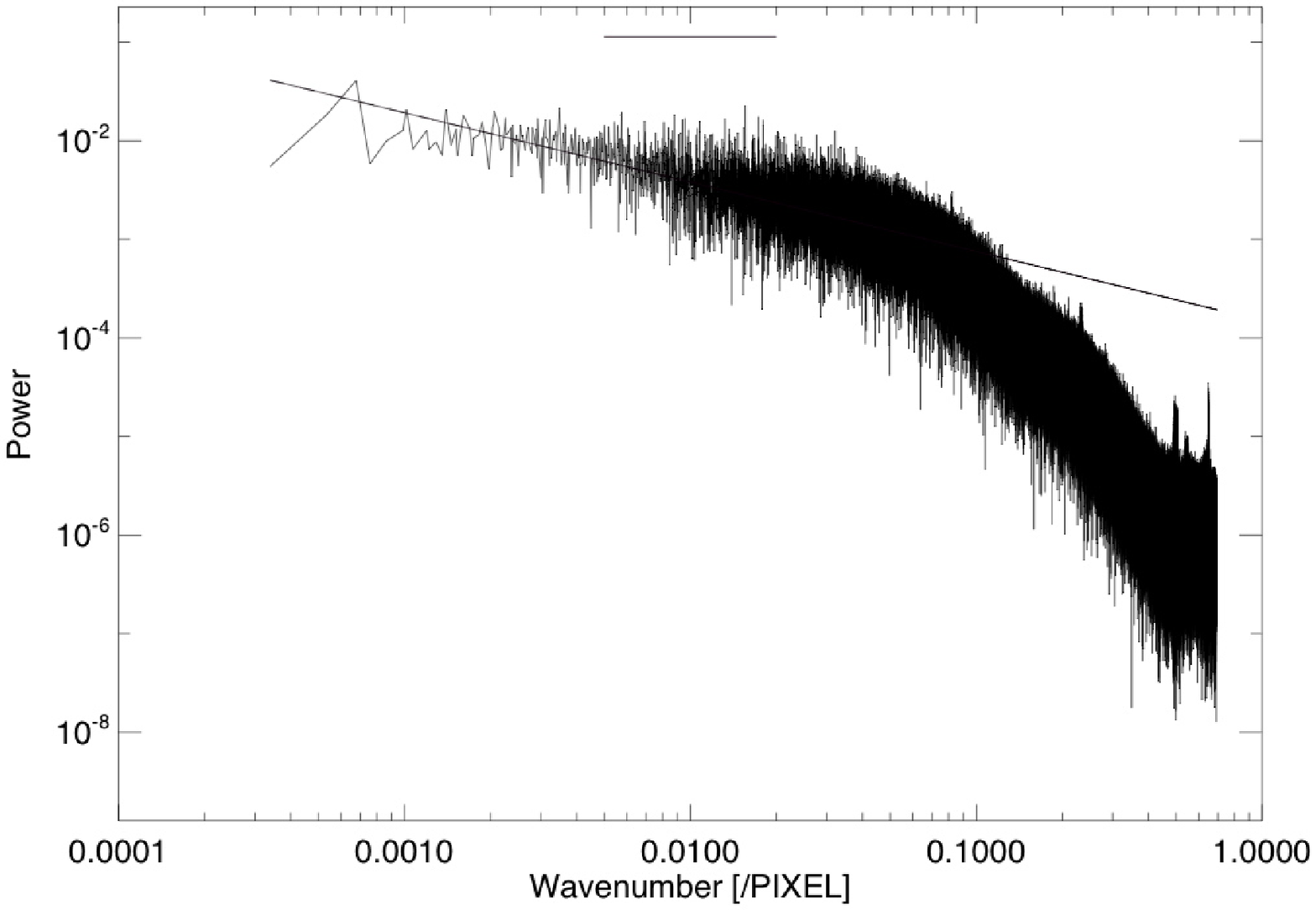,width=8.8cm}
}
\caption{{\it Top-left}: The original standard pipeline processed L18W image of the NEP-Deep survey adapted from Wada et al. (2008). {\it Top-right}: The original standard pipeline processed L18W image of the NEP-Wide survey adapted from Lee et al. (2008).  {\it bottom left}: Power spectrum of the original NEP-Deep survey L18W image. The power spectrum is rather steep indicating a large amount of structure in the background with significant features. {\it bottom right}: The Power spectrum of the original NEP-Wide survey L18W image. The background of the NEP-Wide survey image appears much flatter without artefacts. The straight line fits represent the fit to the power spectrum of the background over the wave number range defined by the upper horizontal bar.
\label{originalpowerspectrum}}
\end{figure*}

The NEP-Deep survey at  18\,$\mu$m in the IRC-L18W band is constructed
from a total of 87 individual pointed  observations taken between  May 2006 to
August 2007, using the  IRC Astronomical Observing Template (AOT)
designed for deep observations (IRC05), with approximately 2500 second
exposures per IRC filter in all mid-infrared bands. The deep imaging
IRC05 AOT has no explicit dithering built into the AOT operation,
therefore dithering is achieved by layering separate pointed
observations on at least three positions on a given piece of sky (as
exemplified in the {\it middle panel} of Figure \ref{convolvedimages},
see Wada et al. \shortcite{wada08} for details). The NEP-Wide survey
consists of 446 pointed observations with  $\sim$300 second exposures
for each filter. The NEP-Wide survey uses the shallower IRC03 AOT
optimized for large area multi-band mapping with the dithering
included within the AOT. Note that for both surveys, although images
are taken simultaneously in all three IRC channels, the  target area
of sky in the  MIR-L channel is offset from the corresponding area of
sky in the NIR/MIR-S channel by $\sim$20 arcmin. However, the 
spiral tiling pattern 
of the NEP survey allows for a 100$\%$ filling factor for all
channels as the satellite field of view rotates about the NEP over a
period of around six months creating a circular survey area. The
final L18W band images for the NEP-Deep and NEP-Wide have been
previously presented in Wada et al.  \shortcite{wada08} and  Lee et
al.  \shortcite{lee09} respectively and are adapted for this work in Figure \ref{originalpowerspectrum}.

\begin{figure*}
\centering
\centerline{
\psfig{ figure=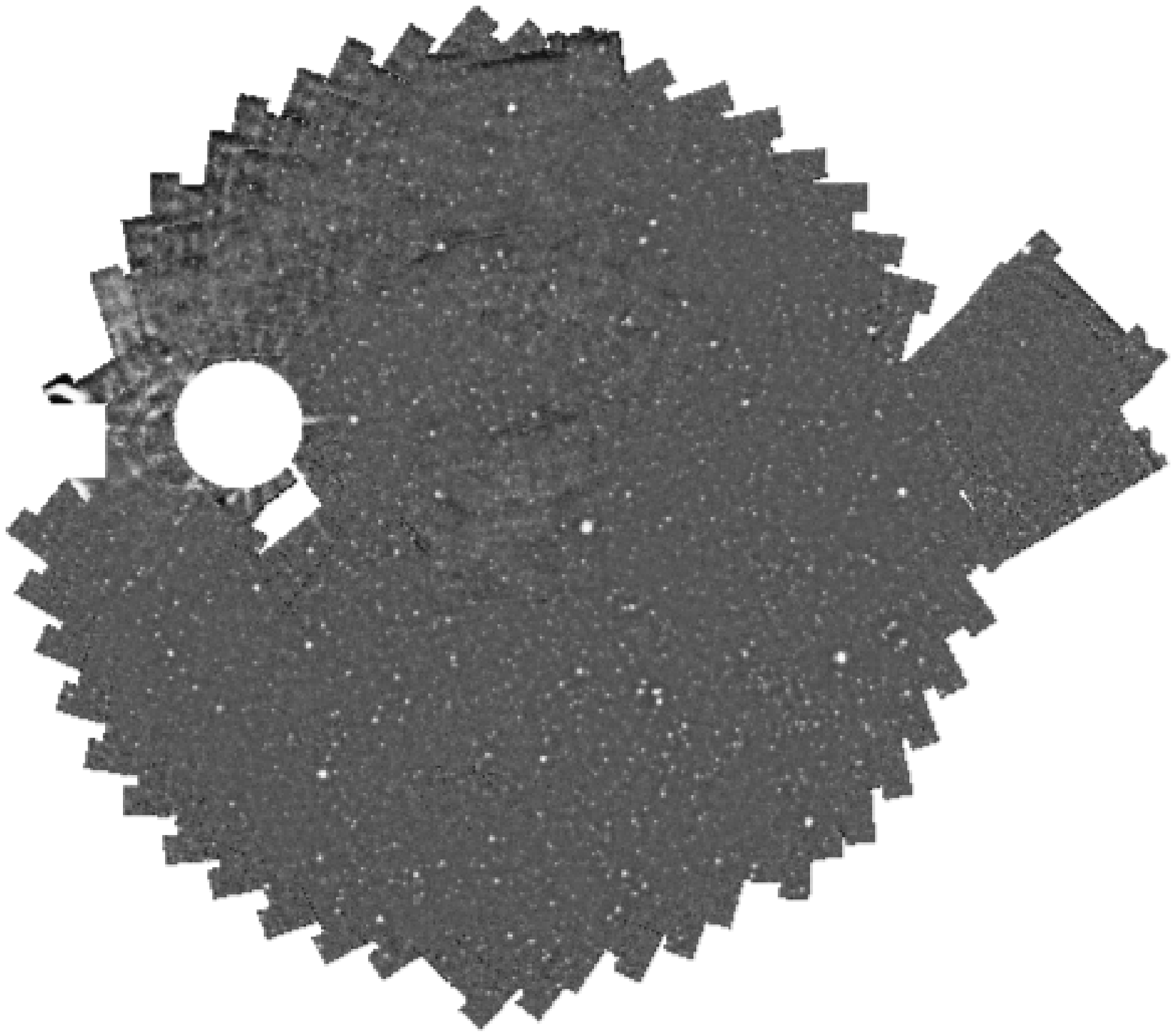,width=8cm}
\psfig{ figure=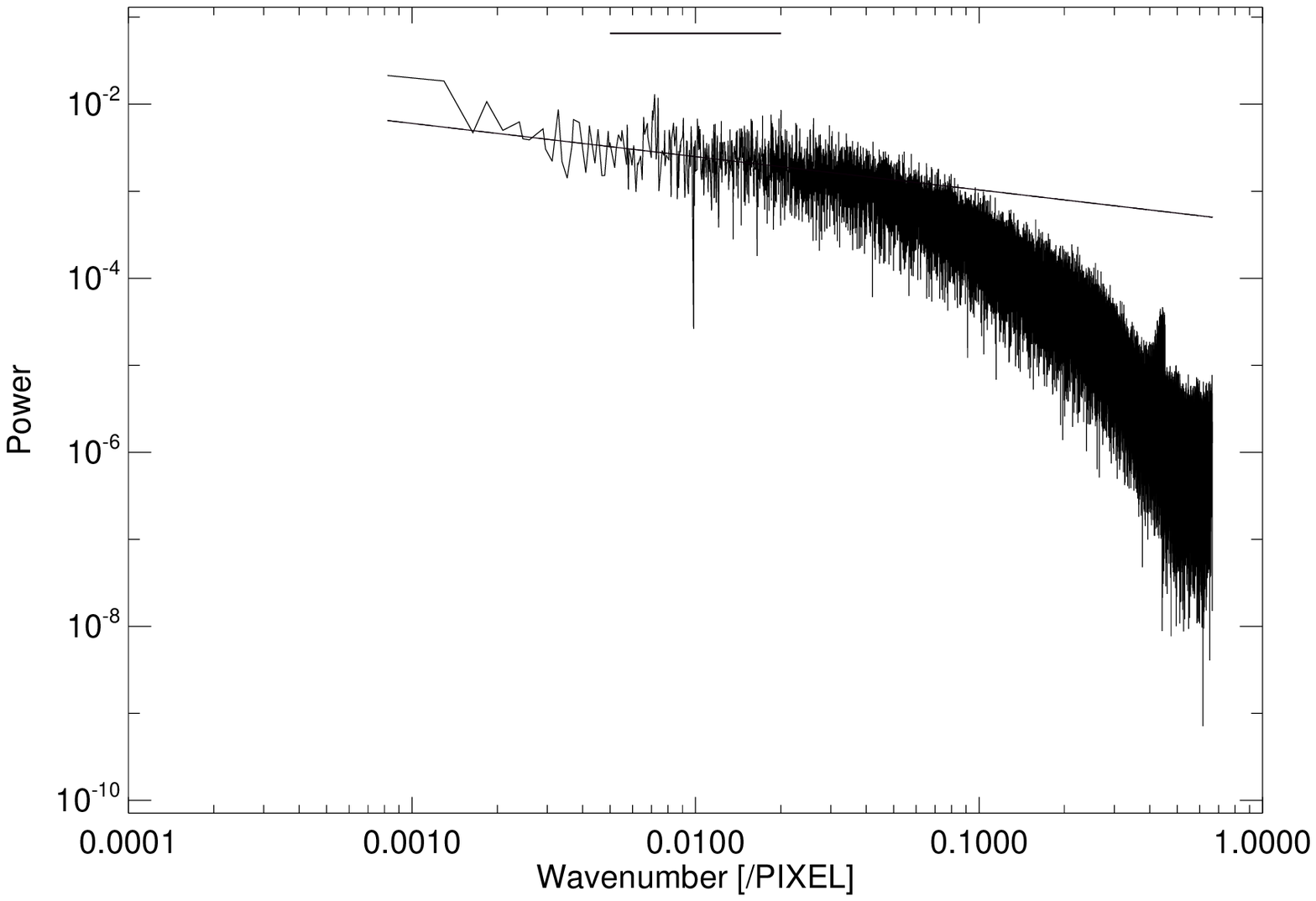,width=8cm}
}
\caption{{\it Top-left}: The masked image of the NEP-Deep survey image, with the mask primarily covering the edges of the image and the area around the bright Cat's Eye Nebula. {\it Top-right}: Window-function-corrected Power Spectrum of the image plus mask region showing a shallower  power spectrum and the removal of the features at large wavenumber. The straight line fits represent the fit to the power spectrum of the background over the wave number range defined by the upper horizontal bar. The window-function-corrected power spectra look like a slightly non-white power spectrum, with a turn over at small scales (large wavenumbers) caused by the PSF, or equivalently caused by the diffraction limit of the telescope. We fit to the power spectrum at a scale before the contribution from PSF. 
\label{mask}}
\end{figure*}  


\subsection{Data Reduction}\label{sec:Reduction}

The data were reduced by the standard IRC imaging pipeline, version
20071017 within the IRAF environment  \cite{lorente07}. The pipeline
processes individual exposure frames (approximately 600 secs
containing $\sim$30 frames per pointing for AOT IRC05) correcting
basic instrumental effects, performing dark subtraction and
flat-fielding using the  {\it AKARI} standard super-dark and flat
frames respectively. It is known that   {\it AKARI} observations
suffer from contamination from diffuse light due to Earth shine in the
Summer months, and although the background of each frame is removed by
subtracting the self-median image, the condition still results in the
rejection of frames, especially in the MIR-L bands (15, 18,
24\,$\mu$m). The individual frames are co-added to make the image for an
individual pointed observation using the average pixel values rather
than the median to optimize the signal-to-noise. Cosmic ray events,
etc, are also removed at this stage using an outlier clipping
technique. 
The output from the standard pipeline is thus a set of individual 10
$\times$ 10 arcmin pointing tiles that are then mosaiced using the
the publicly available software
SWarp\footnote{http://terapix.iap.fr/rubrique.php?id rubrique=49}, 
after attaching the world coordinate system (WCS) on each image. Note
that the   {\it AKARI} pipeline calculates astrometry by matching
bright stars in the image with corresponding sources in the 2MASS star
catalogue. However for observations taken with the IRC MIR-L channel,
the number of bright stars are often insufficient to correctly
calculate the shift and rotation of the image to match  the 2MASS
data, therefore the data taken simultaneously in the   {\it AKARI}
MIR-S/NIR channel is used to project the correct astrometry onto the
MIR-L images.

\section{Post Processing of the NEP L18W Band image}\label{sec:processing}

\subsection{Image Processing}\label{sec:imageprocessing}
The original NEP Deep L18W image exhibits some structure where either
data was lost or corrupted, and where the mosaicing was less than
perfect especially at the edges and centre of the image. In addition
the Cat's Eye Nebula (NGC 6543) at 17$^h$58$^m$33.4$^s$
+66$^\circ$37$^m$59$^s$  with a 25\,$\mu$m flux density of 110\,Jy \cite{moshir90}
dominates the left portion of the image affecting the surrounding area.
After-images from detector hysteresis are apparent in a clockwise
direction from NGC 6543.
Such structure may considerably hinder
effective source extraction, therefore in order to visualize the
degree of structure in the NEP images we measure the power spectrum of
both the NEP-Deep and NEP-Wide images. The original images and associated power spectra are shown in
Figure \ref{originalpowerspectrum} for the NEP-Deep {\it left} and NEP-Wide
{\it right} images. Optimally we would prefer as flat a  background as
possible over a large range in wavenumber, as can be seen to be the
case for the NEP-Wide survey. However,  the power spectrum of the
NEP-Deep image shows significant structure with a steeper spectrum. 

\begin{figure*}
\centering
\centerline{
\psfig{ figure=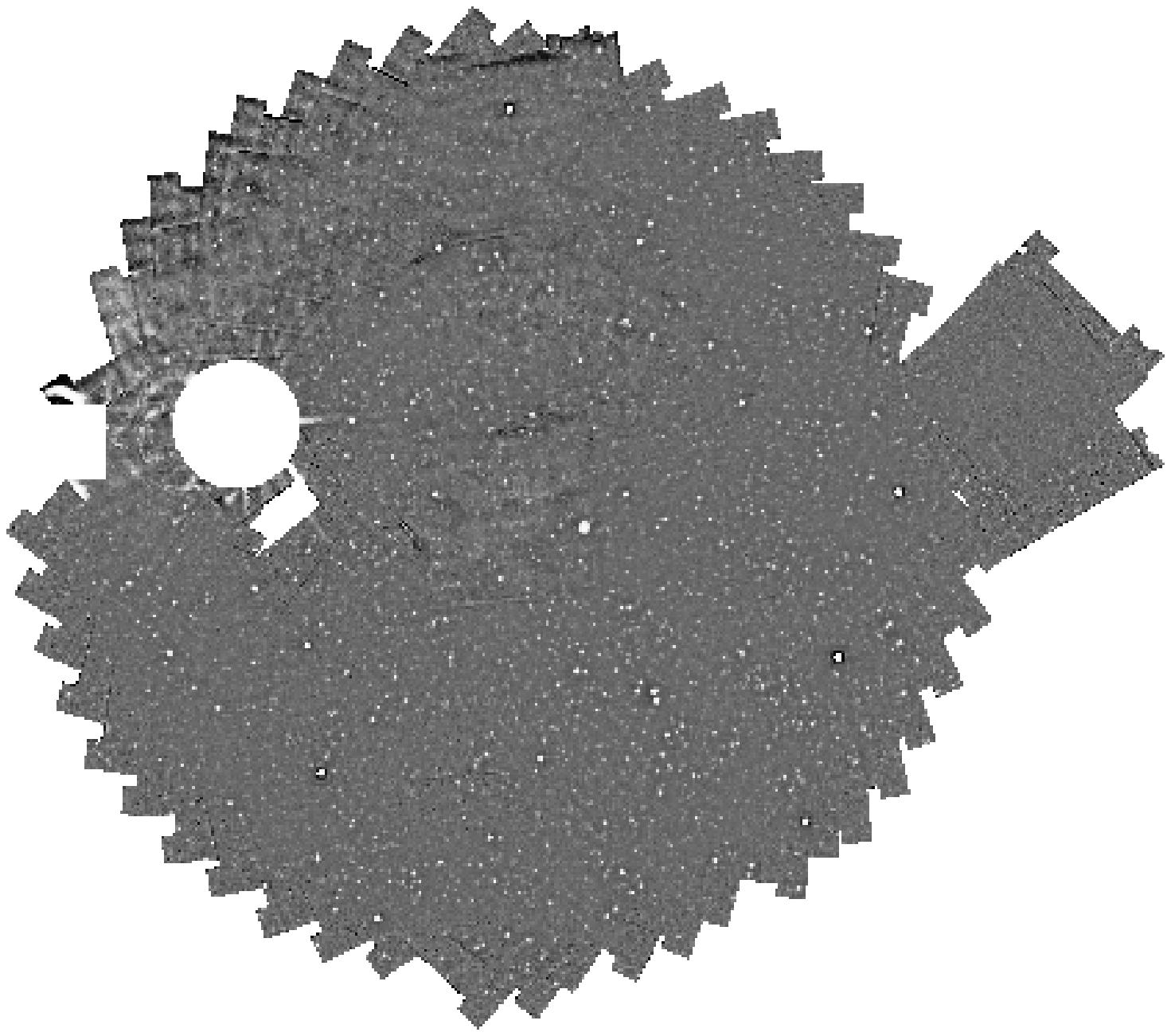,width=9cm}
\psfig{ figure=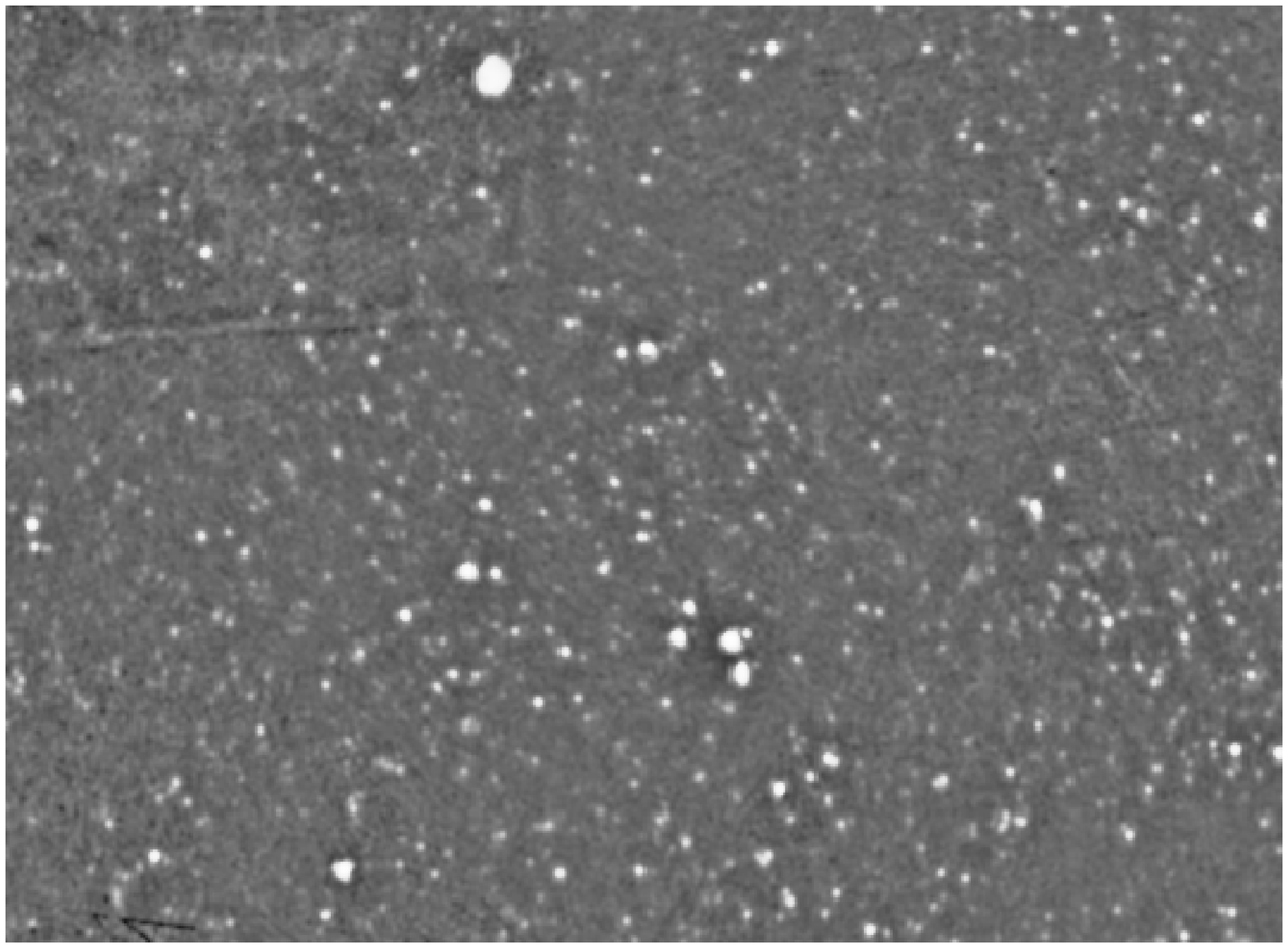,width=7cm}
}
\centerline{
\psfig{ figure=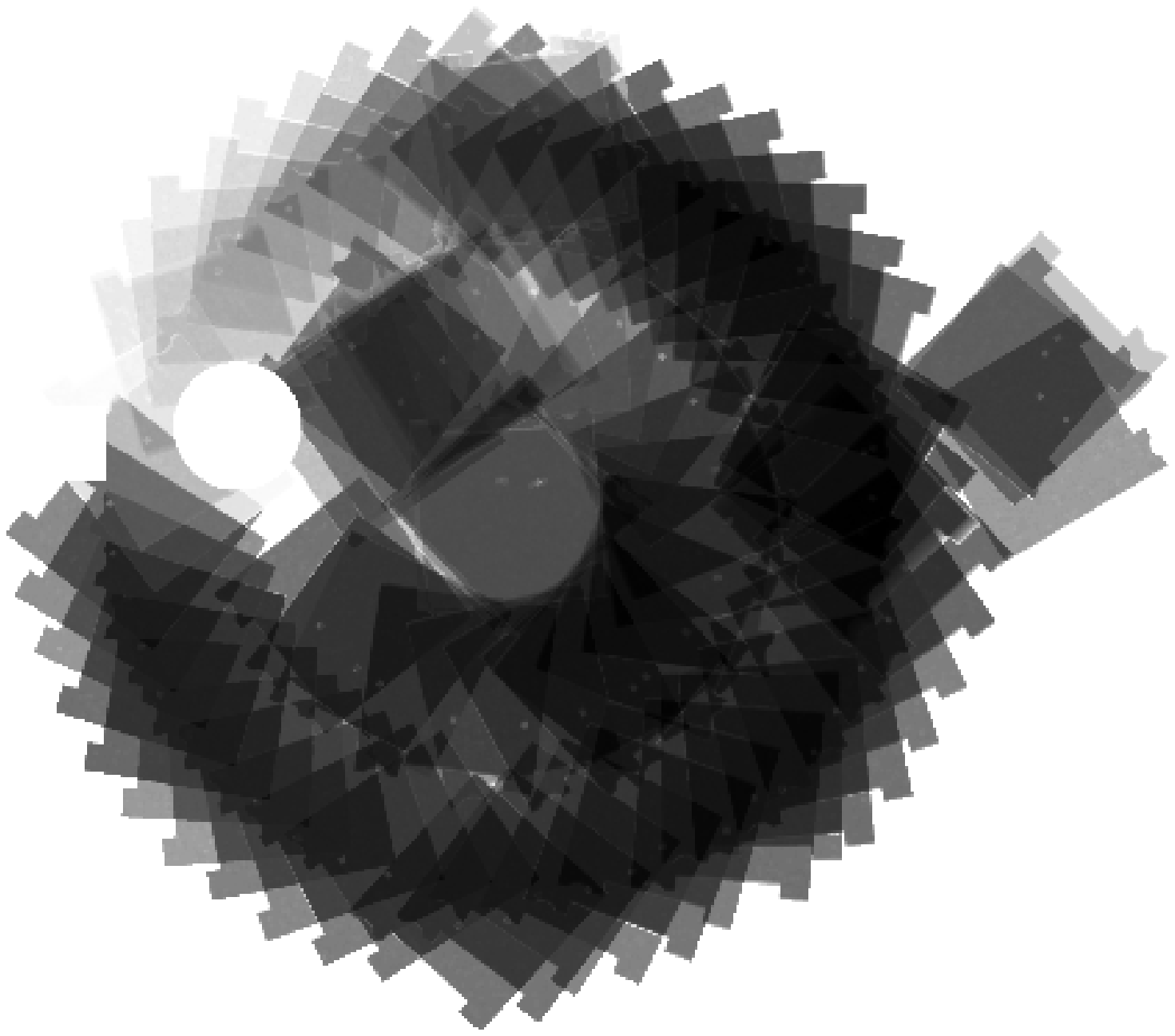,width=9cm}
\psfig{ figure=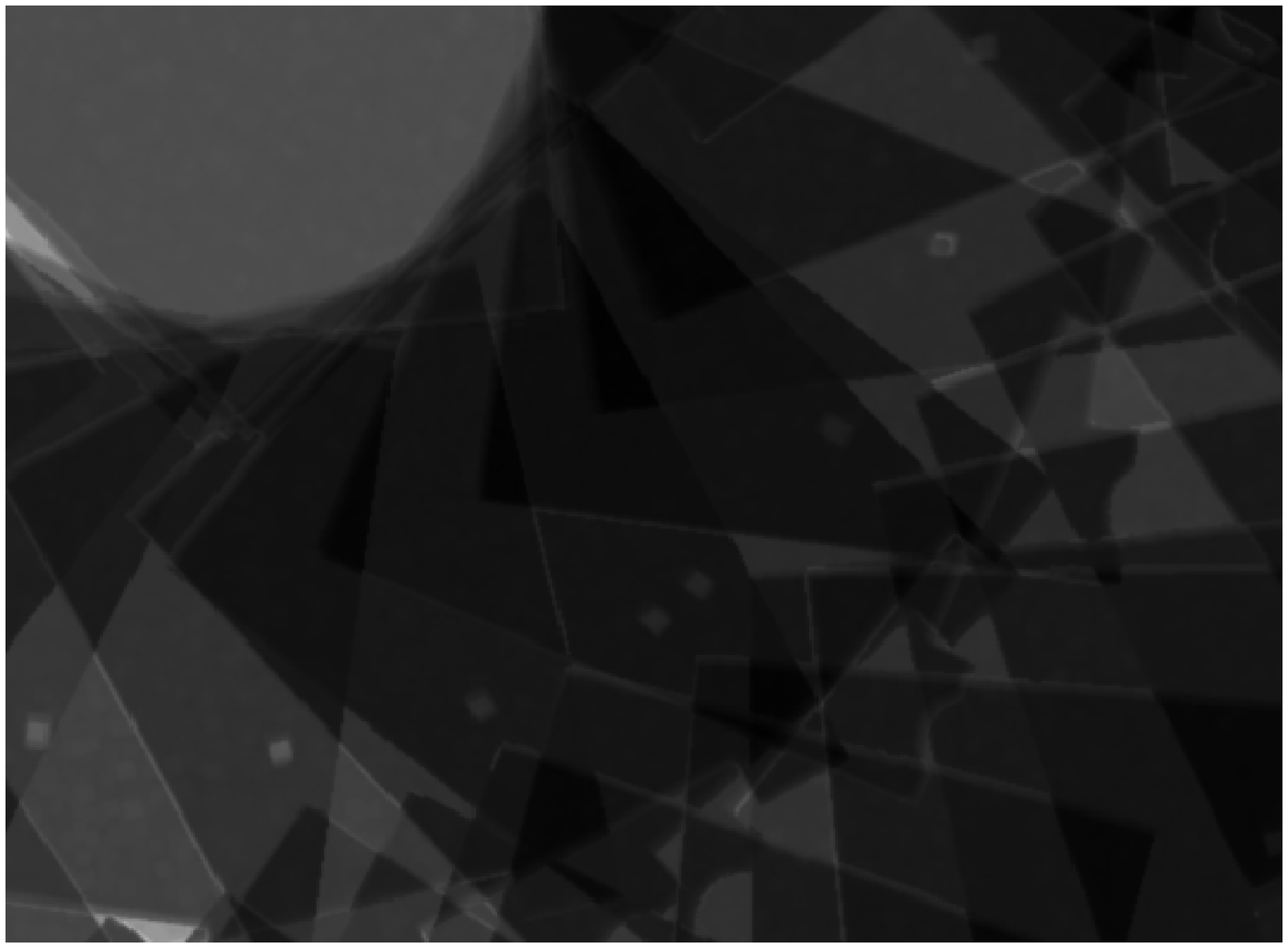,width=7cm}
}
\centerline{
\psfig{ figure=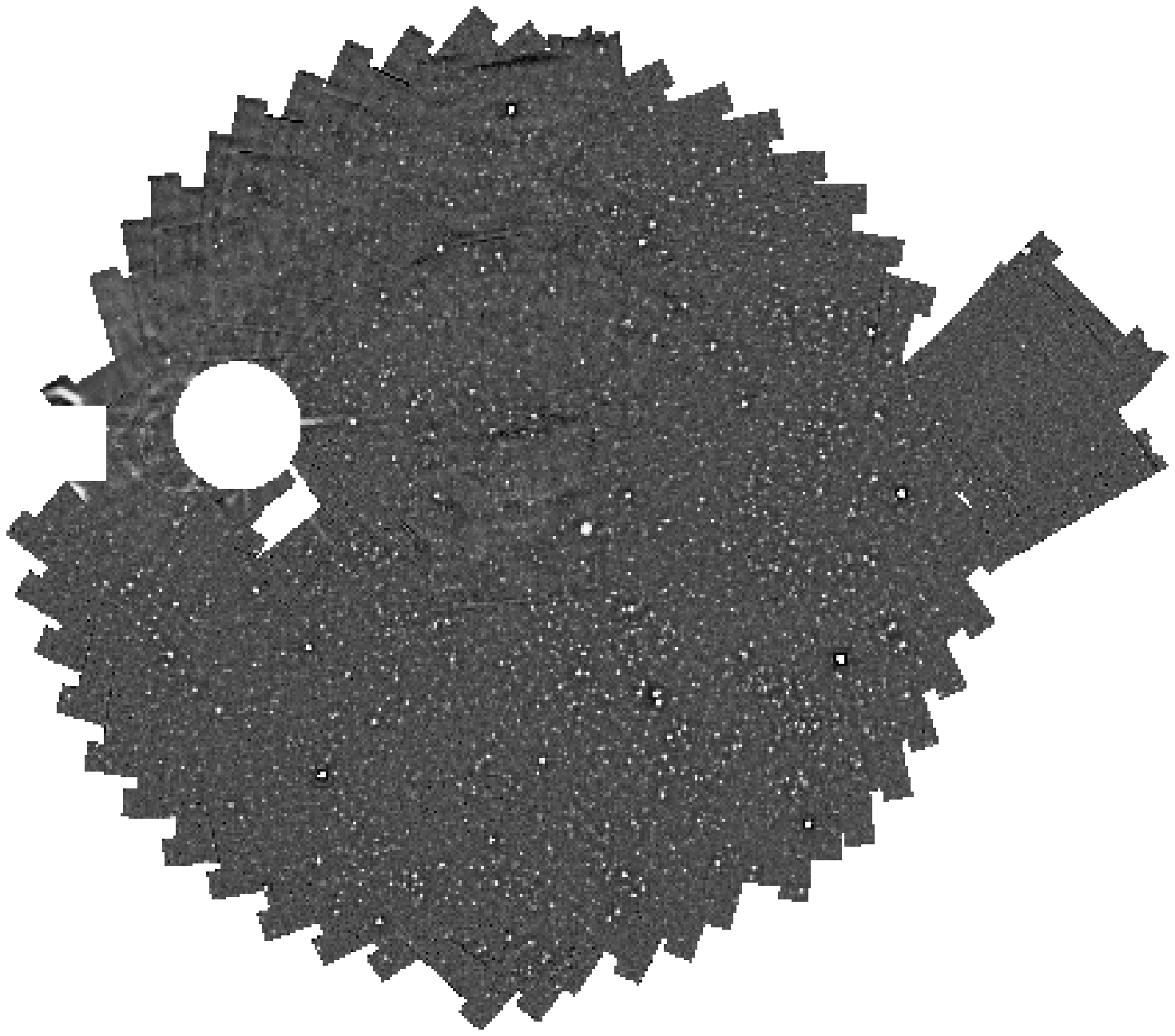,width=9cm}
\psfig{ figure=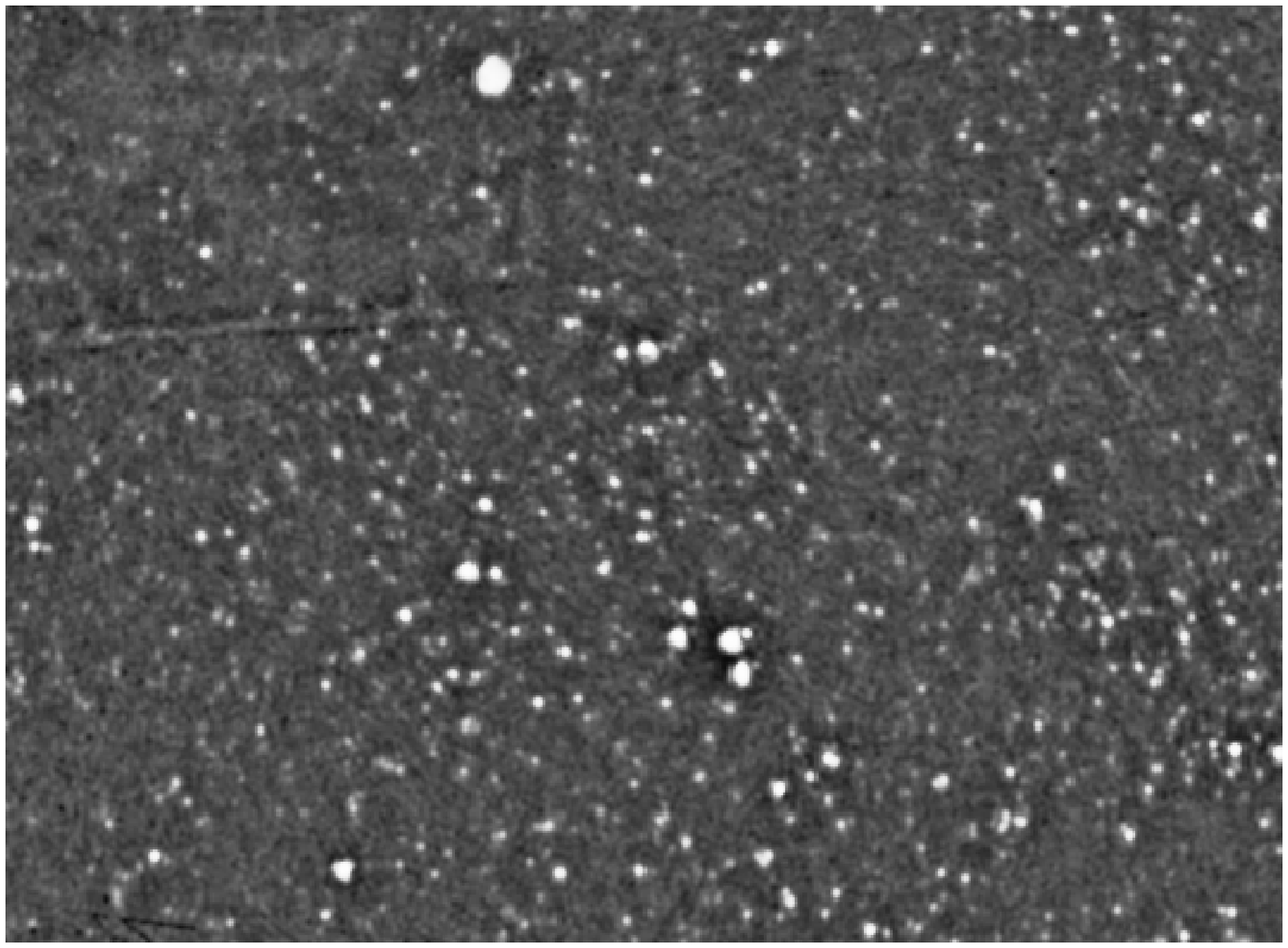,width=7cm}
}
\caption{Result of the convolution of the original L18W image with the
  optimal noise-weighted (minimum $\chi ^{2}$) point source filter
  using the PSF supplied from the Wiener filtering process. From top
  to bottom, the PSF convolved image, noise map and signal to noise
  image. The {\it right panels} show magnified portions of the same
  images for detail. 
\label{convolvedimages}}
\end{figure*}  

We assume that much of the structure in the NEP-Deep image is
contributed by the influence of the Cat's Eye Nebula and the edges of
the image. We therefore mask the edges of the image and create a
circular mask around the Cat's Eye Nebula region. In Figure \ref {mask}  we show the NEP-Deep image with mask cut out  with the corresponding power spectrum. From the Figure it can indeed be seen that most
of the structure seen originates from the suspected regions and we
therefore multiply the image by this binary mask. The resulting power
spectrum, although not perfect, is significantly smoother and
shallower, more closely resembling the NEP-Wide power spectrum. Note that the drop off in the power spectrum to higher wave numbers is not in fact caused by additional structure in the map but rather due to the convolution of the map with the PSF (since the Fourier Transform of a convolution is equivalent to a multiplication, we see a drop off at larger wavenumbers). We therefore fit to the power spectrum at a scale before the contribution from PSF.

  In order to recover the point sources from the structured background
  we apply a technique known as Wiener filtering developed originally
  for CMB decontamination to 
remove point sources  \cite{vio02}. In the spirit of "one persons'
signal being anothers' noise" and vice-versa,  this technique was
reversed to filter IRAS sources from images with a  structured
background \cite{hwang07b}. Conceptually simple, the
Wiener filtering presumes that for a given point spread function
$F$, a Gaussian background with power spectrum $P_{k}$ with wave
number $k$, and Fourier transforms denoted by a tildes, $\sim$, the
Fourier transform of the minimum-variance point source filter is
simply given by  $\widetilde{F}_{\rm Wiener} \propto
\frac{\widetilde{F}}{P_{k}} $. The Wiener filter algorithm returns
the optimal signal-to-noise point source filter from the power spectrum and filters
the background in Fourier space producing a point source filtered
image. The optimal filter superficially resembles a Mexican Hat wavelet
with the central positive peak slightly narrower than the point spread
function.  The Wiener filtering algorithm is also performed on the
NEP-Wide image producing  point source filtered images and more importantly, optimal point source detection smoothing kernels for both the NEP-Deep and NEP-Wide images.\\

\subsection{Source Extraction}\label{sec:extraction}

Our source extraction is carried out by convolving our image+mask in
Figure \ref{mask} with the optimal point source filter produced from
the Wiener filtering algorithm. We use the method of Serjeant et
al. \shortcite{serjeant03} which was originally applied to
submillimetre SCUBA surveys. This method operates on the premise that
for a raw image $I$ with noise level, $N$ and weight map $W=1/N^{2}$,
the best-fit optimal noise-weighted minimum $\chi ^{2}$ point source
flux density, $S$, anywhere in the map, assuming a given point spread function
$F$, is 
given by $S= (I \times W)\bigotimes F) / (W\bigotimes
F^{2})$, where $\bigotimes$ denotes a convolution.
The errors on the fluxes are given by $\Delta S =(W\bigotimes
F^{2})^{-1/2}$. Bad pixels in the image can easily be dealt with by
assigning arbitrarily high noise values to them, removing the need for
a separate interpolation stage or replacement of bad pixels with local
mean values. 
The rebinning in the final image mosaicing stage nevertheless induces 
correlations between the pixels prior to the convolution, 
which results in an under-estimated noise value in the point source
filtered map. We therefore re-scaled the noise values of the 
filtered map to ensure the 
core of the signal-to-noise histogram had a unit variance. 
The results of the convolution of the original L18W image with the
optimal point source filter are shown in Figure \ref{convolvedimages}
with from {\it top} to {\it bottom}, the PSF convolved image, noise
map and signal to noise image. These resulting maps give the best-fit
flux and best-fit errors at any given point in the maps. Also shown in
Figure \ref{convolvedimages} are magnified selections of the same
sub-region in each map for closer inspection. One can see in some
instances the dark rings around some sources produced by the optimal
PSF.

Sources are then extracted by  a thresholding technique performed on
the signal to noise image where by sources are 
selected 
from
diminishing thresholds from a S/N of 10$\sigma$ down to 5$\sigma$ by
selecting a source as a number of connected pixels above a given S/N
threshold. In this technique, double peak sources with differing S/N
occupying connected pixel positions in the image can be distinguished
by successively lower S/N thresholds allowing effective selection and
extraction of individual peaks as individual sources
(e.g. Mortier et al.  ~\shortcite{mortier05}). 
 The best-fit flux (in raw counts) for each extracted source is given
 by the value in the convolved image at the extracted source position
 and the corresponding errors from the values in the noise map.

\section{Photometry}\label{sec:photometry}

\subsection{Calibration of the minimum variance method}\label{sec:calibration}

The final products processed by the standard IRC pipeline are produced
as images calibrated in instrumental units (ADU). The {\it AKARI} IRC Instrument manual
\cite{lorente07} provides conversion factors from the raw instrument
ADU units to Jy.
For the L18W band the conversion factor of ADU/s to $\mu$Jy for point sources is 1ADU =
1.148\,$\mu$Jy. This conversion factor is based on aperture photometry
of standard stars assuming for the L18W band a 7.5 pixel radius for
the flux measurement and a surrounding annulus from 7.5-12.5 pixels
for the sky measurement \cite{tanabe08}. Clearly, this conversion
factor is not applicable for our optimum PSF convolved minimum
variance algorithm and indeed using this conversion factor yields
meaningless results for the source fluxes. 

\begin{table*}
\caption{Raw counts to flux density conversion factors for photometry methods calculated from {\it AKARI} calibration stars. Note that a comparison of the actual scaling factors is meaningless and it is the percentage errors that are significant. The Wiener filtered PSF providing the lowest error margin.}
\begin{tabular}{@{}lllllll}
Calibration Star & Flux Density& Wiener Filter PSF & APER & SExtractor \\
                         &  (milli-Jansky) & \multicolumn{3}{c}{raw counts to micro-Jansky conversion factor} \\
\hline
BD+66 1073 & 13     & 6.329 $\pm$0.55$\%$  & 1.328 $\pm$1.41$\%$  & 1.369 $\pm$1.12$\%$  \\
KF01T4     & 8.16         & 6.397 $\pm$0.86$\%$  & 1.218 $\pm$1.61$\%$  & 1.333 $\pm$1.56$\%$ \\
BD+66 1060 & 26.4  & 6.427 $\pm$0.24$\%$  & 1.104 $\pm$0.69$\%$  & 1.117 $\pm$0.38$\%$  \\
\hline
\end{tabular}
\label{tab:calibration}
\end{table*}

Therefore, we  returned to the original calibration stars in order to calculate the appropriate
conversion factors for our photometry method. The full list of calibration stars used for the {\it AKARI} flux calibration is given in Tanabe et al. \shortcite{tanabe08}. Although, each calibration star was observed in dedicated pointing observations,  we note that three of these calibration stars actually lie within the NEP-Wide field itself, namely BD+66 1073, KF01T4 \& BD+66 1060. Therefore for comparison, we reprocessed the raw data from these individual calibration star observations using exactly the same method previously applied to the NEP L18W images in order to  produce identically reduced images. We then compared the results with the measured signal from the same calibration stars within the NEP image itself.

We use the optimal PSF calculated from the Wiener Filtering, from the NEP image and convolve it with the standard
star image (i.e. we perform exactly the same procedure on the standard
stars as we have done on the NEP image). The thresholding is then made
for all sources on the standard star image. We use a Digital Sky
Survey (DSS\footnote{http://archive.stsci.edu/dss/}) image to locate
our standard stars in the IRC image and measure the raw counts given
by the minimum variance source extraction.

Comparing the raw counts with the expected calibration flux  presented in Tanabe et al. \shortcite{tanabe08}, using the
 spectral models of Cohen et al. \shortcite{cohen96}, \shortcite{cohen99}, \shortcite{cohen03a}, \shortcite{cohen03b},
should provide a new conversion factor applicable to our minimum variance source extraction method. 
However, we note that the measured signal in instrumental units varies for each standard star depending on whether it was measured in the individual image or the NEP image itself by up to 10$\%$. This can be understood in part by the fact that the  optimal filter was explicitly derived for the NEP field. 

In addition to our optimum filter specifically applicable to the NEP images, we have also carried out photometry on the
three standard stars in both the individual calibration images and the NEP image with two other methods for illustration. We have used both the SExtractor algorithm \cite{bertin96} and the APER routine inside IDL\footnote{Interactive  data Language: http://www.exelisvis.com/ProductsServices/IDL.aspx}  to carry out aperture photometry  utilizing the same parameters as those for the  standard star calibration and the initial data reduction of the NEP region carried out by Wada et al. \shortcite{wada08}, Takagi et al.  \shortcite{takagi12b} (i.e. 7.5 pixel radius aperture for the flux and a surrounding 5 pixel annulus for the sky. Again we note, to lesser degree a discrepancy in the signal measured for the stars depending on the field (individual calibration field or the NEP field) of the order of  6$\%$ and 2$\%$ for the SExtractor and APER routines respectively. Although these discrepancies are consistent with the $\pm$6$\%$ absolute flux calibration quoted by Tanabe et al. \shortcite{tanabe08}, we conservatively conclude to calculate our conversion factors using the measurements of the standard stars within the NEP image itself, as shown in   Table \ref{tab:calibration}.

From Table  \ref{tab:calibration} we see that in general
the results for the individual calibration stars are in reasonable agreement
within any given method to a few percent. However, the results from different methods
can differ significantly. Examining the errors also show us that in
general we are obtaining a better signal to noise for  any given
measurement for the minimum variance methods compared to the aperture
photometry methods. 

In Table \ref{tab:conversion} we calculate the appropriate conversion
factor for each of our photometry methods by calculating the mean of
the conversion factors from each calibration star. For our minimum
variance source extraction assuming the optimal filter we arrive at
a  conversion factor of 1ADU = 6.38\,$\mu$Jy.  From the Table it can
be seen that the appropriate conversion factor is sensitive to the
method of photometry.  Note that the values obtained using the APER and
SExtractor aperture photometry methods are within 5-6$\%$ of the
value derived for the {\it AKARI} IDUM = 1.146 $\pm$4.6$\%$, also
derived via aperture photometry.

A closer comparison with the aperture photometry methods again
emphasizes the improvement in signal to noise gained by using the
minimum variance point source extraction algorithm, with typical
errors being $\sim$2 times better than the aperture photometry results.

\begin{table}
\caption{Final conversion factors from raw counts to micro-Janskys for the photometry methods tested in Table \ref{tab:calibration}. Note that a comparison of the actual scaling factors is meaningless and it is the percentage errors that are significant. The Wiener filtered PSF providing the lowest error margin of the three methods.}
\centering
\begin{tabular}{@{}ll}
Photometry Method & Counts to $\mu$Jy conversion factor \\
\hline
Wiener PSF    & 6.384 $\pm$1.06$\%$     \\
APER                              & 1.217 $\pm$2.34$\%$     \\
SExtractor                       & 1.273 $\pm$1.96$\%$     \\
\hline
\end{tabular}
\label{tab:conversion}
\end{table}

\begin{figure}
\centering
\centerline{
\psfig{ figure=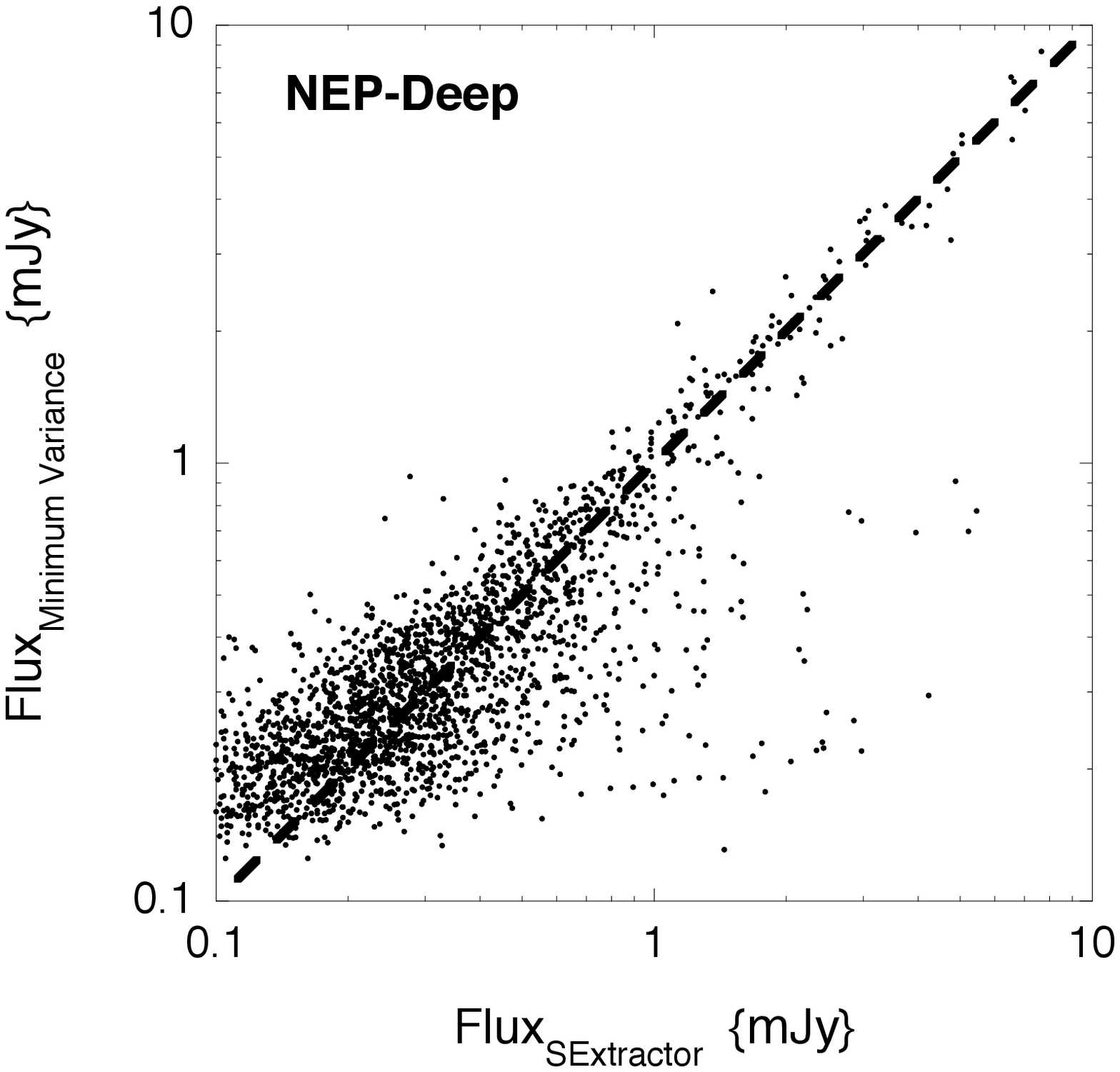,width=8cm}
}
\centerline{
\psfig{ figure=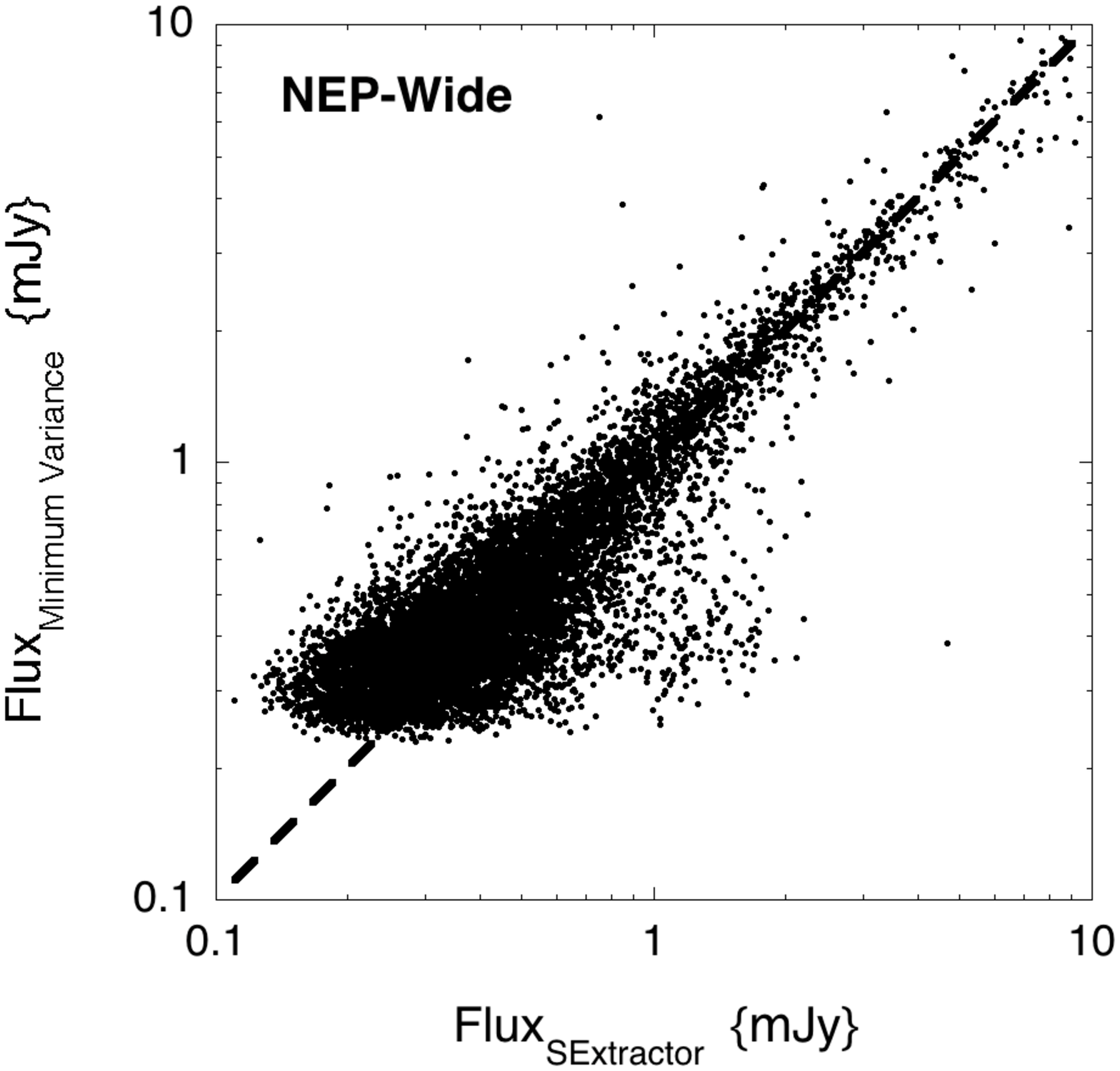,width=8cm}
}
\caption{A comparison of the photometry results given by the  minimum variance source extraction method and the flux densities from the results of Wada et al. (2008) using SExtractor. All sources are marked as being $>$5$\sigma$ in both catalogues.
\label{fluxfluxcompare}}
\end{figure}  

\begin{figure}
\centering
\centerline{
\psfig{ figure=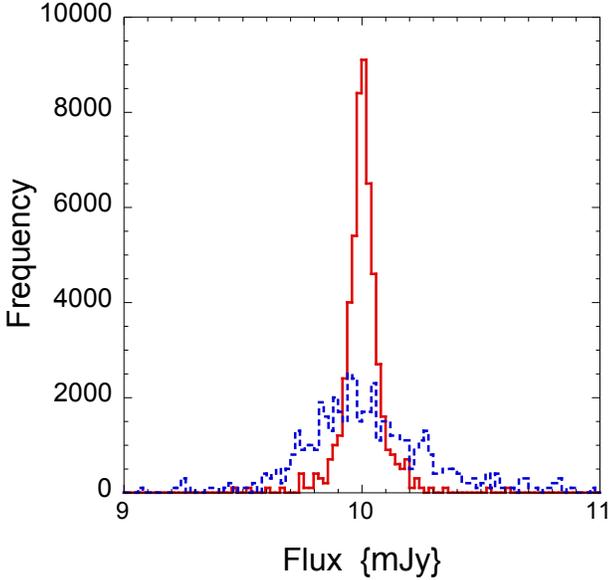,width=8cm}
}
\caption{Results for photometry from Monte Carlo simulations injecting artificial sources of flux density 10mJy into the original NEP Deep image. Results are shown for photometry using the PSF fitting from the minimum variance filter method (tall narrow line in red) compared to conventional aperture photometry (lower line in blue). Although the extracted fluxes agree, note the significant dispersion of the fluxes measured via aperture photometry about the mean, compared to the minimum variance filter method.
\label{simulations}}
\end{figure}  

\subsection{Verification of the minimum variance method}\label{sec:verification}

\begin{figure*}
\centering
\centerline{
\psfig{ figure=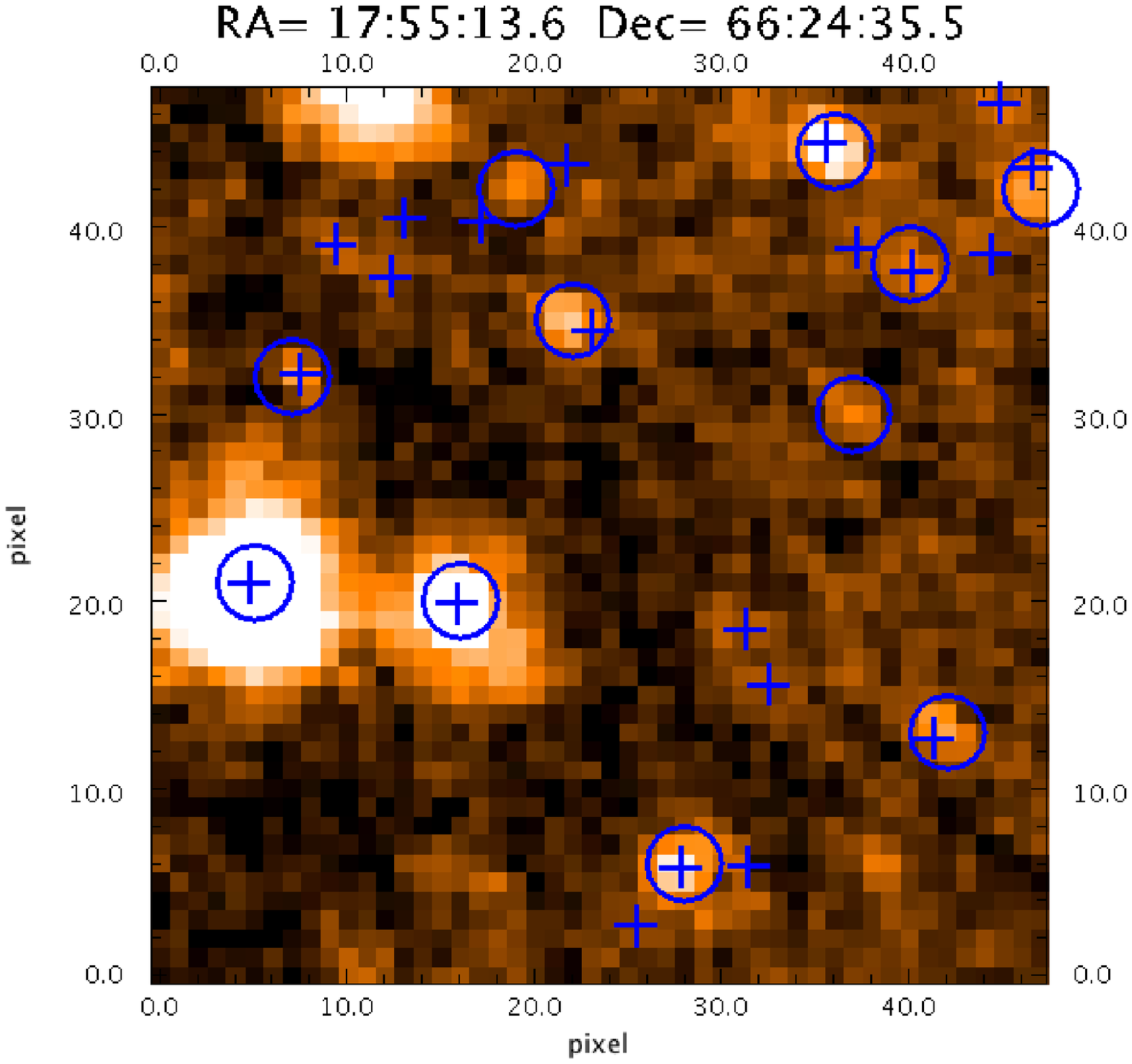,width=7cm}
\psfig{ figure=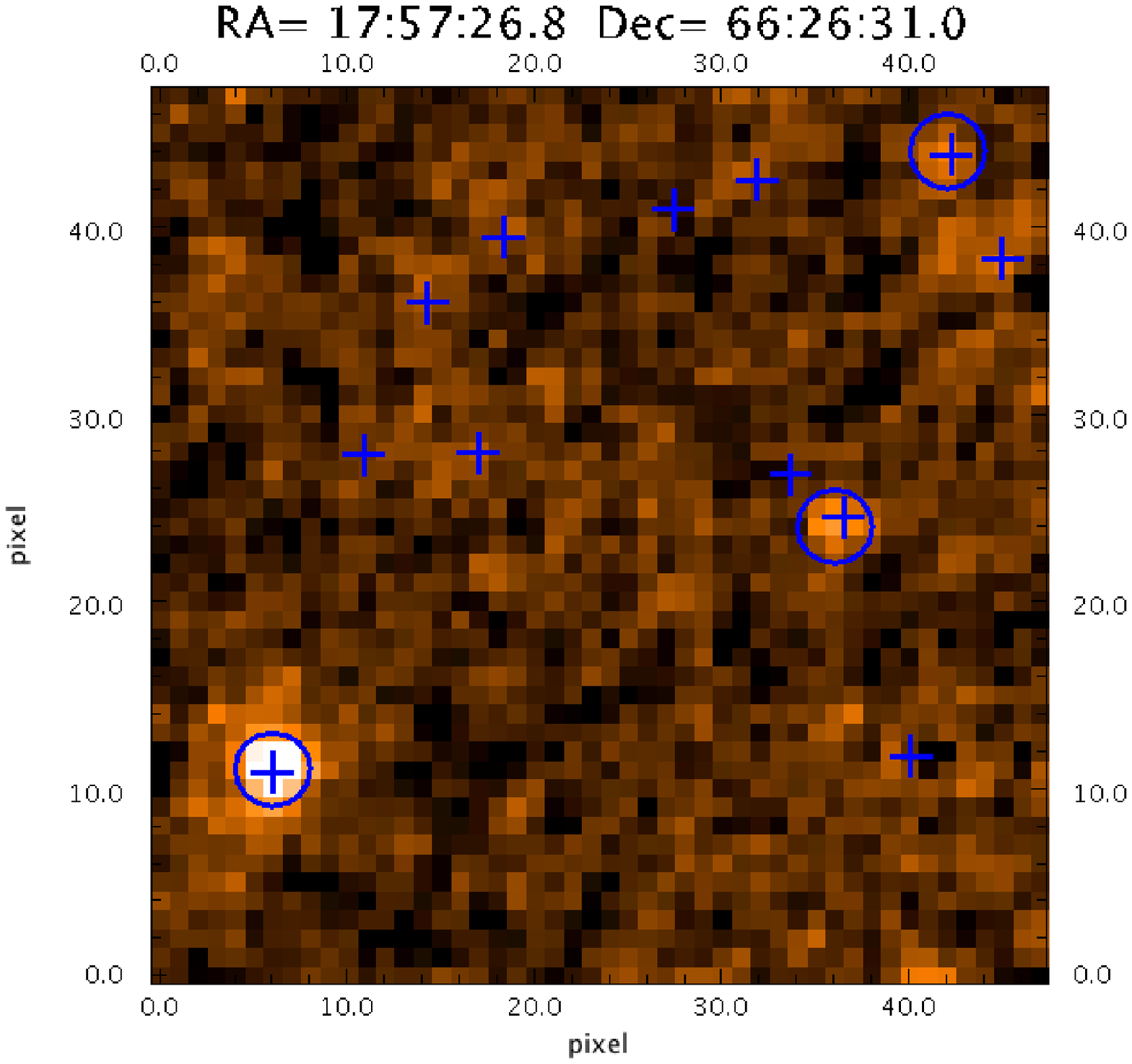,width=7cm}
}
\centerline{
\psfig{ figure=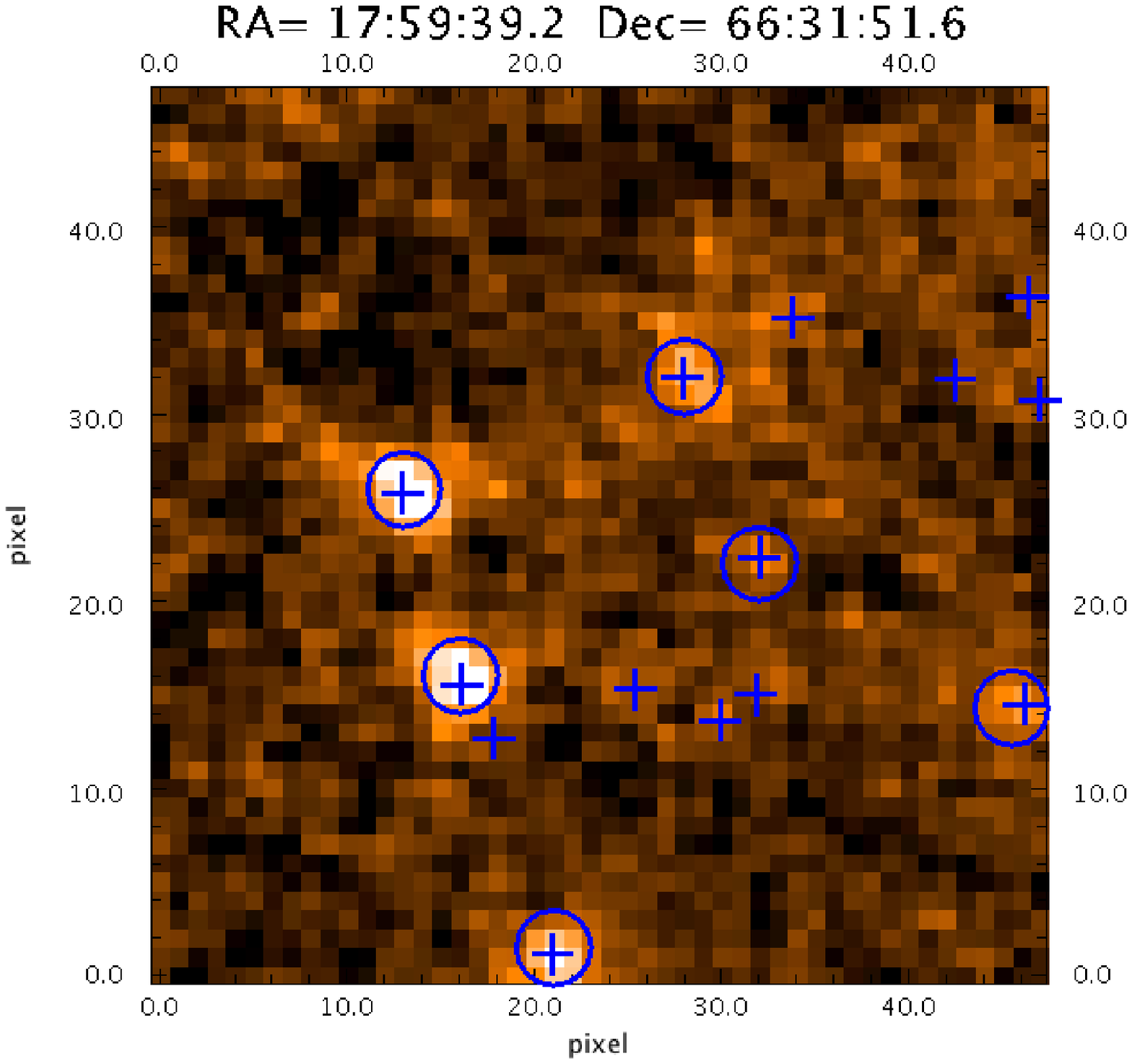,width=7cm}
\psfig{ figure=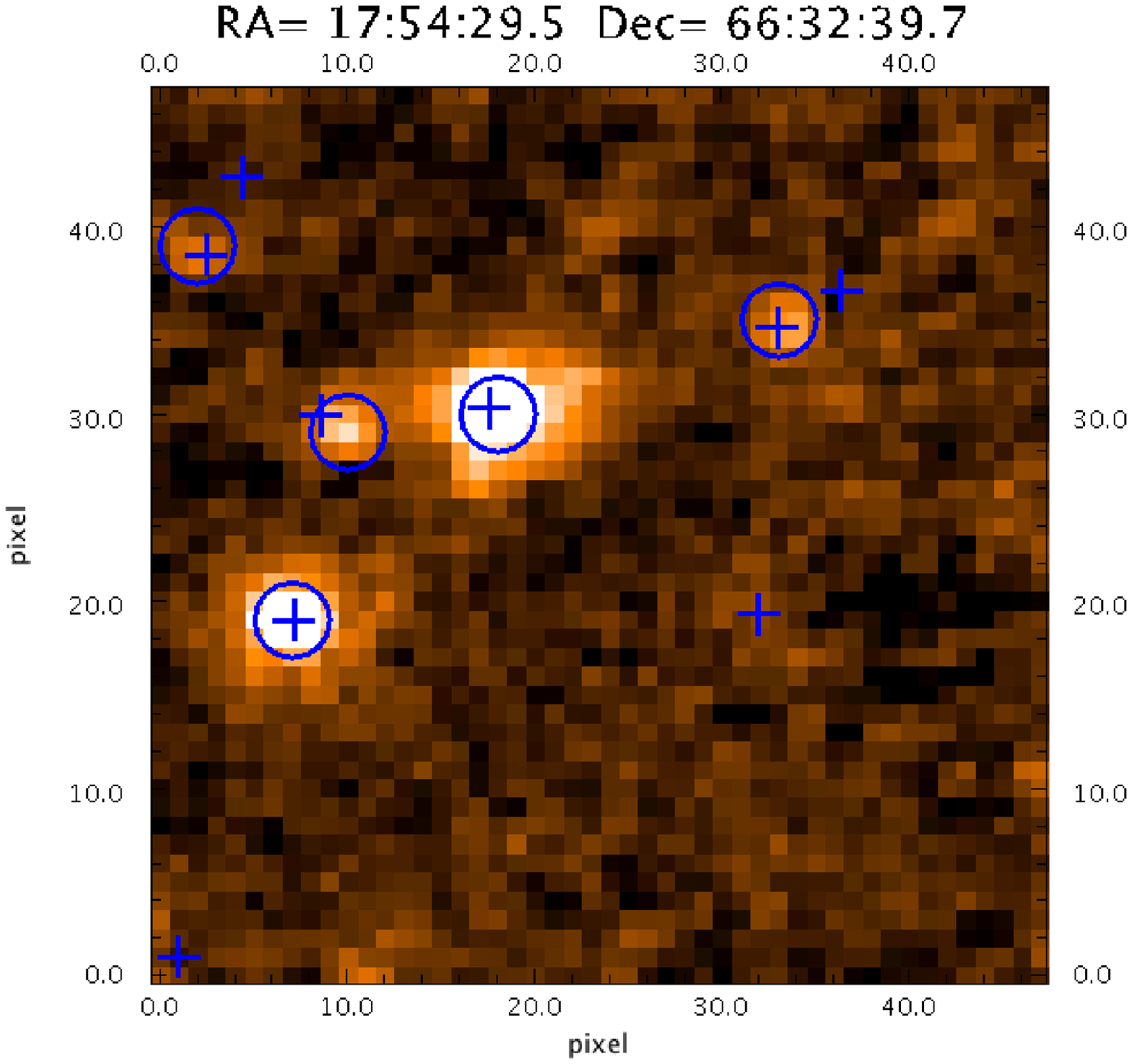,width=7cm}
}
\caption{Postage stamp images of high signal to noise sources in the L18W NEP-Deep map (1$\times$1 arcmin). The {\it crosses} are sources in the mass processed catalogue of Wada et al. (2008). {\it open circles} are $>$5$\sigma$ sources from the  minimum variance source extraction method. Many examples are seen where faint sources in the catalogue of Wada et al. (2008) do not appear to have any reliable counterpart in the map. There are also cases where the minimum variance source extraction method finds reliable sources not included in the catalogue of Wada et al..
\label{postage}}
\end{figure*}

As a test of our photometry methods we have taken our catalogue
produced using the minimum variance source extraction method and
compared it with the catalogues produced by Wada et al.
\shortcite{wada08}. This L18W catalogue was produced as a product of
the general mass processing of the NEP-Deep data set using SExtractor
to identify the sources. Aperture photometry was performed on the
extracted sources using a 7.5 pixel (18$\arcsec$) radius for the
photometry and a sky annulus of 7.5 --12.5 pixels.  Wada et al. \shortcite{wada08} quote a 5$\sigma$ sensitivity of 0.121mJy and an 80$\%$ completeness limit of 0.2mJy in the L18W band.
We  cross-correlated the two catalogues using the publicly available TOPCAT
software\footnote{http://www.star.bris.ac.uk/~mbt/topcat/} assuming a
separation for matching sources of 5$\arcsec$ which corresponds to
approximately 2 pixels on the image, achieving $\sim$4000 unique
matches. In Figure \ref{fluxfluxcompare} ({\it top-panel}) we show the resulting flux density
comparison for the sources found in both catalogues 
(all sources are marked as being $>$5$\sigma$ in both catalogues). We find a reasonable
agreement between both source extraction and photometry methods at
least down to the mJy level, although a slight underestimation of the SExtractor fluxes in the Wada et al.
\shortcite{wada08} estimates is visible. At fainter fluxes we observe a large
scatter between the two catalogue fluxes. Note that from postage stamps shown in Figure \ref{postage}, their are examples from the catalogue of Wada et al. \shortcite{wada08}  where two or more entries are associated around a single brighter source resulting in fainter flux density estimates compared to the single source extracted by the minimum variance method.

In addition we also make the same comparison with the NEP-Wide catalogue of Kim et al. \shortcite{kim12} finding a total of 9800 matches using the same search criteria. The results are shown in the {\it bottom-panel} of Figure \ref{fluxfluxcompare} and confirm the consistency of the photometry derived from the minimum variance filter PSF.

{\bf 
We also investigated any effect on the photometry due to the differences in the PSF derived from the Weiner filtering on the calibration stars since the wide {\it AKARI} filters may result in broader PSFs for red sources than for blue sources like stars. Note that a PSF modeller for {\it AKARI} not available, however we have used the Spitzer STINYTIM \footnote{http://irsa.ipac.caltech.edu/data/SPITZER/docs/ \\ dataanalysistools/tools/contributed/general/stinytim/} software for computing PSF models to investigate the effects of different spectral energy distributions on the PSF. Using STINYTIM, assuming a FWHM for the {\it AKARI}  IRC L18W band the software predicts a negligible difference ($1.2\%$) between the simulated PSFs of a star sampled on the RayleighÐJeans tail and a $\nu F _{\nu}$ = constant spectrum typical of galaxies in the mid-infrared. 
}

As a further test of our method and photometry we carried out a series of Monte-Carlo simulations. Artificial sources were injected into the original NEP-Deep image in Figure \ref{originalpowerspectrum} ({\it top-left-panel}) at regular spaced intervals at predetermined positions so as not to be contaminated by real existing sources. Source extraction and aperture photometry using the same method described in  Wada et al.\shortcite{wada08} was then carried out. Following this, the simulated map was convolved with our optimal point source PSF derived from the match filter technique and again the sources were extracted and photometry made on the convolved map. This process was repeated 1000 times for any given flux density and the flux density of the simulated sources was varied between 1 and 100mJy. In Figure  \ref{simulations} we show a representative example from the simulation of artificial sources with truth flux density of 10mJy. Both the matched filter and aperture photometry successfully recover the artificial source flux density, however the fluxes as measured by the matched filter are much more tightly constrained around the truth flux density of 10mJy compared to the fluxes measured by aperture photometry which exhibit a much broader spread of values.

\begin{figure*}
\centering
\centerline{
\psfig{ figure=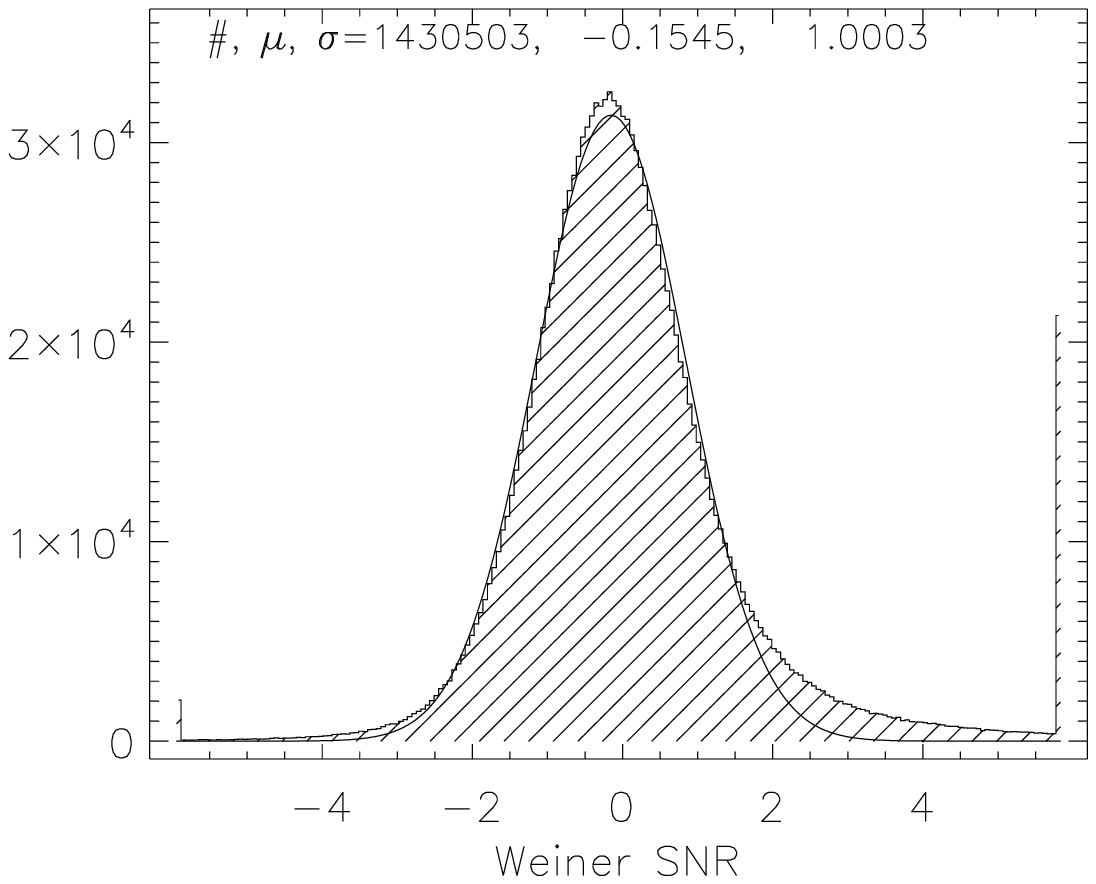,height=5cm}
\psfig{ figure=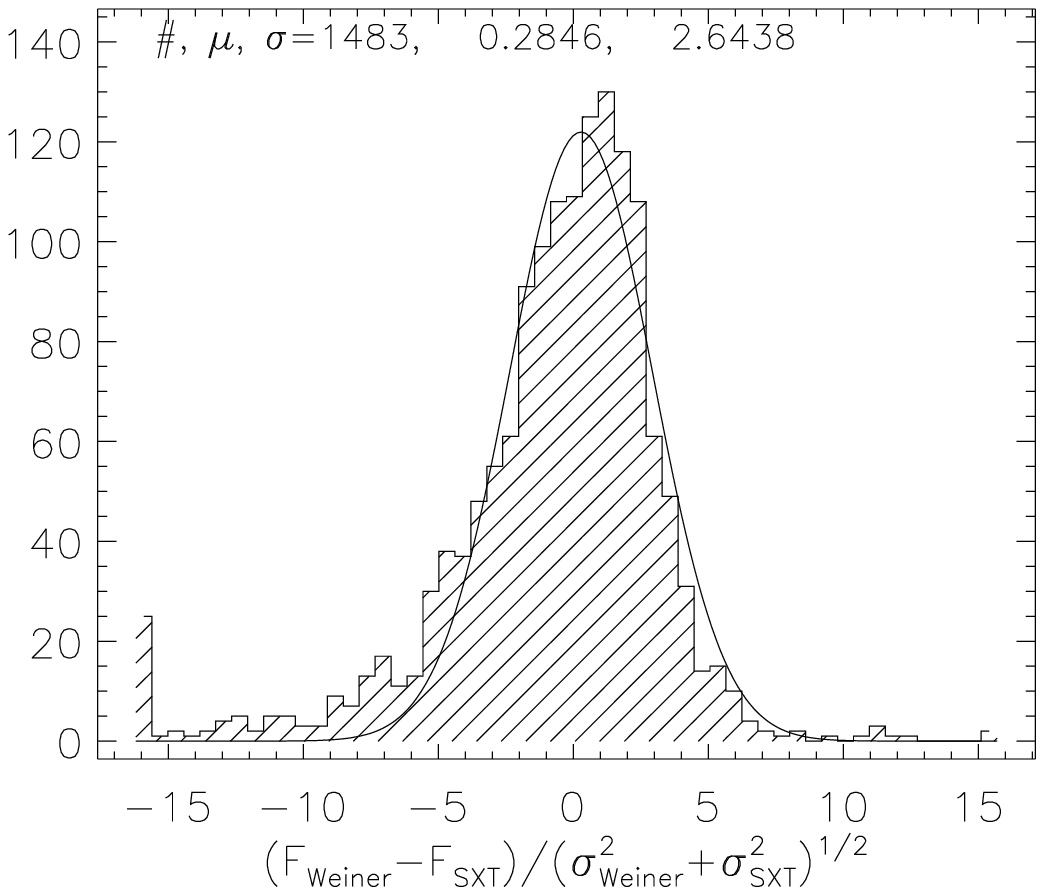,height=5cm}
\psfig{ figure=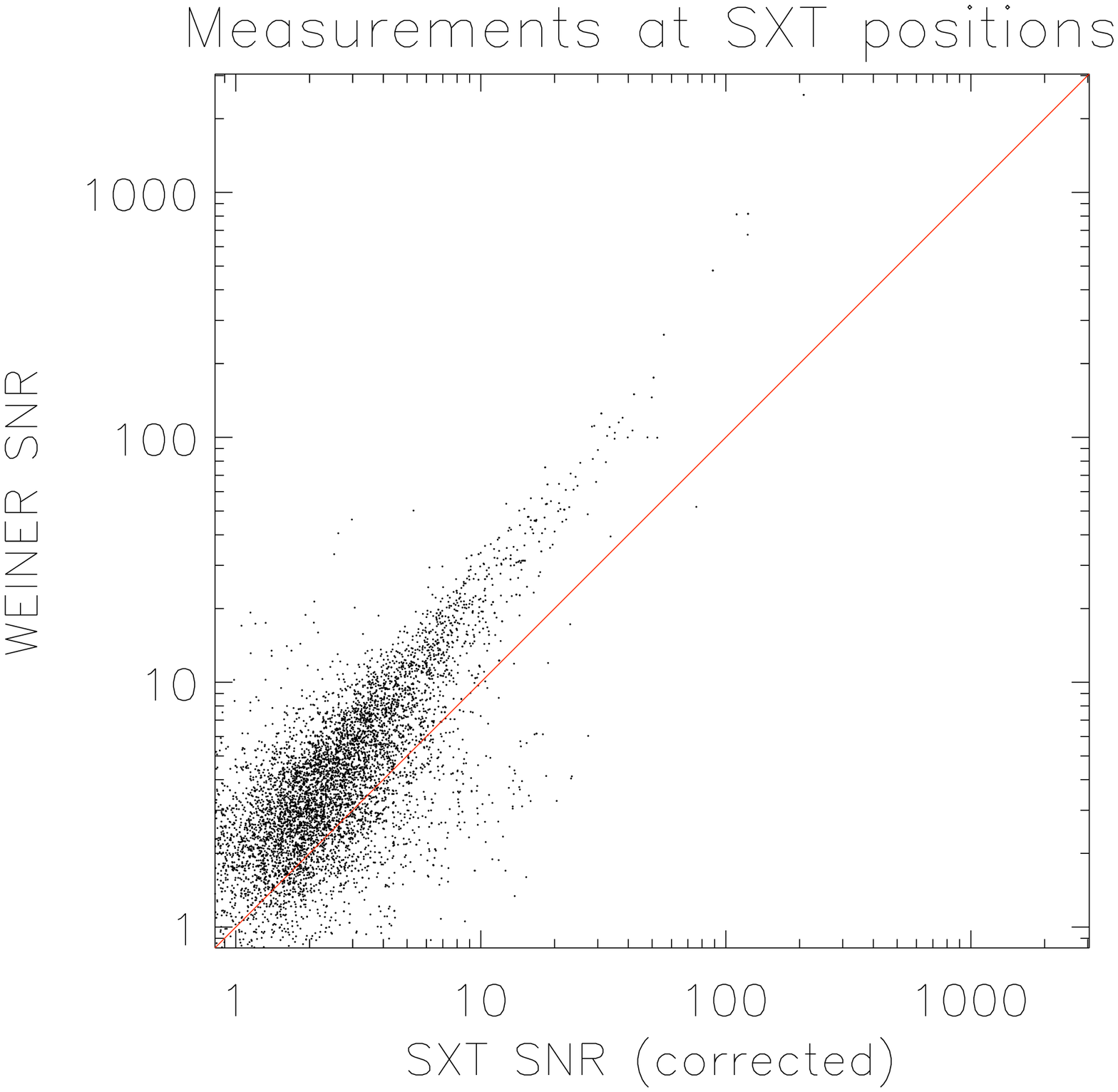,height=5cm}
}
\caption{{\it Left}: The noise histogram estimated from the convolved image noise map of the NEP-Deep image used for the minimum variance source extraction method. Note the unit variance implying that the noise is correctly measured. {\it Middle}: A test for excess noise in the method implemented by Wada et al. (2008). The quantity of the differences in the fluxes of the  minimum variance and SExtractor methods, divided by the square root of the sum of the squares of the noise is plotted for the two methods. Two independent  measurements would give $\sigma$=1 if the noise estimates were correct or sigma$>$1 if the noise has been  under-estimated in the latter method. {\it Right}: The resulting signal to noise plots of our two comparative methods after correction for the  under-estimate of the noise in the SExtracted sources.
\label{noise}}
\end{figure*}  

 In order to estimate the accuracy of the errors we created a noise
 histogram from our optimal PSF convolved noise map. The noise
 histogram is shown in the {\it left panel} of Figure \ref{noise}. The
 signal to noise in the noise histogram has a unit variance implying
 that the noise is reliably estimated. 
Recall that in the case of our minimum variance source extraction method, the noise was rescaled
after convolution. 
 Note that the process of mosaicing the individual pointings together also
 involves a degree of resampling,    inducing correlations between the signal in neighbouring pixels, 
 and this raises a concern about the  accuracy of the errors in the mass processed catalogue. 

Therefore, as a test for excess noise in the method implemented by
Wada et al. \shortcite{wada08} we plot in the {\it middle panel} of
Figure  \ref{noise} the quantity ($S_{\rm Wiener} - S_{\rm SXT}) /
\sqrt{\sigma^{2}_{\rm Wiener} + \sigma^{2}_{\rm SXT}}$, where the $S$
and $\sigma$ values correspond to the Wiener (minimum variance) and SXT
(SExtractor) methods respectively.  

Two independent  measurements would give $\sigma$=1 if the noise
estimates were correct or sigma$>$1 if the noise has been
under-estimated in the latter method. It can be seen that although
from Figure \ref{fluxfluxcompare} our measured flux densities reassuringly
correlate strongly with the SExtractor result, from the {\it middle
  panel} in Figure \ref{noise}, the noise in the Wada et al. mass processed
catalogue has been underestimated 
{\bf 
(i.e. we see a standard deviation of 2.6438 rather than unity).  
}

We believe that the noise in this case has been
underestimated because the source extraction in the mass produced
catalogue assumes the pixels in the image are statistically
independent. They are not independent, because the earlier mosaicing procedure introduces
covariances between the pixels. We therefore renormalise the signal to noise histogram to ensure a variance of unity to correct this underestimate of the noise in the original image as measured by SExtractor.

Then  in the {\it Right panel} of Figure \ref{noise}  we plot the resulting signal to noise plots of our two
comparative methods . We find that the  minimum variance source extraction gives signal-to-noise (S/N) values
higher by  about a factor of 1.9 compared to the SExtractor catalogue. This result is also consistent with
our reanalysis of the calibration stars where we find a similar difference in the S/N between the the flux densities measured by the minimum variance method and aperture photometry with SExtractor of $\sim$1.5.

To investigate the quality of the two catalogues we have also made postage stamp images of high signal to noise sources. Figure \ref{postage} shows $>$ 5$\sigma$ selected sources from the catalogue
produced from the minimum variance source extraction catalogues with sources from our catalogue as  {\it open circles} . The {\it crosses} are sources in the mass processed catalogue of Wada et al. \shortcite{wada08} .  It can be seen that for the brighter sources, in most (but not all) cases the catalogues match, however there also seem to be many cases where the mass processed catalogue marks a source where seemingly there is only noise. 

In addition, on visual inspection there are also many examples where the original catalogue has two (or more) entries  around a single brighter source in the image, corresponding to a single source entry in our catalogue. 
This issue also results in fainter spurious flux density estimates in the original catalogue.

Our comparison of our source flux densities, confidence in our noise estimates
and visual checking of our maps therefore lead us to believe that our
minimum variance source extraction method is indeed a more robust
estimate of both source position and flux density.

\section{Source Counts}\label{sec:Source Counts}

\subsection{Raw Source Counts}\label{sec:raw}

\begin{figure}
\centering
\centerline{
\psfig{ figure=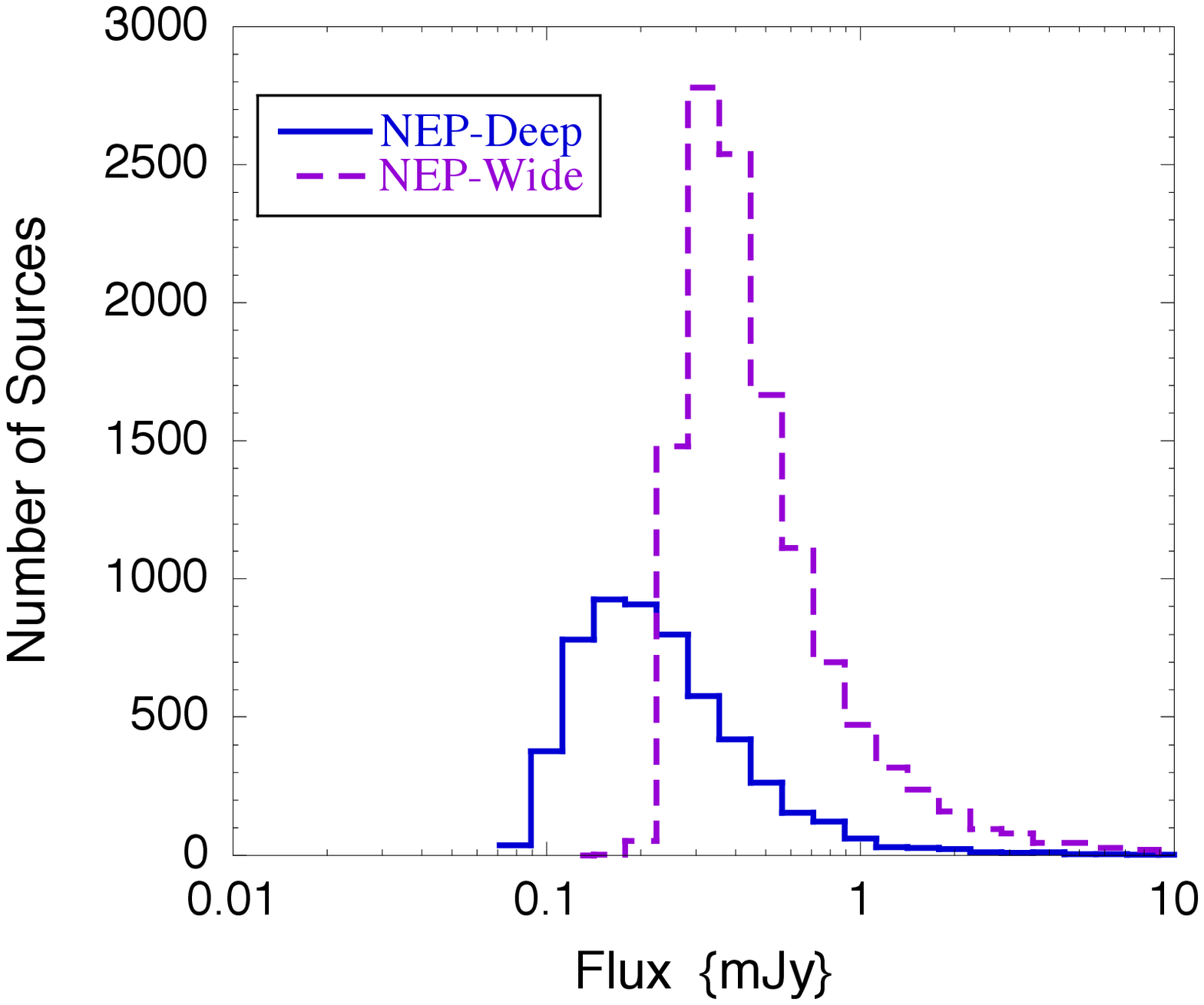,width=7cm}
}
\centerline{
\psfig{ figure= 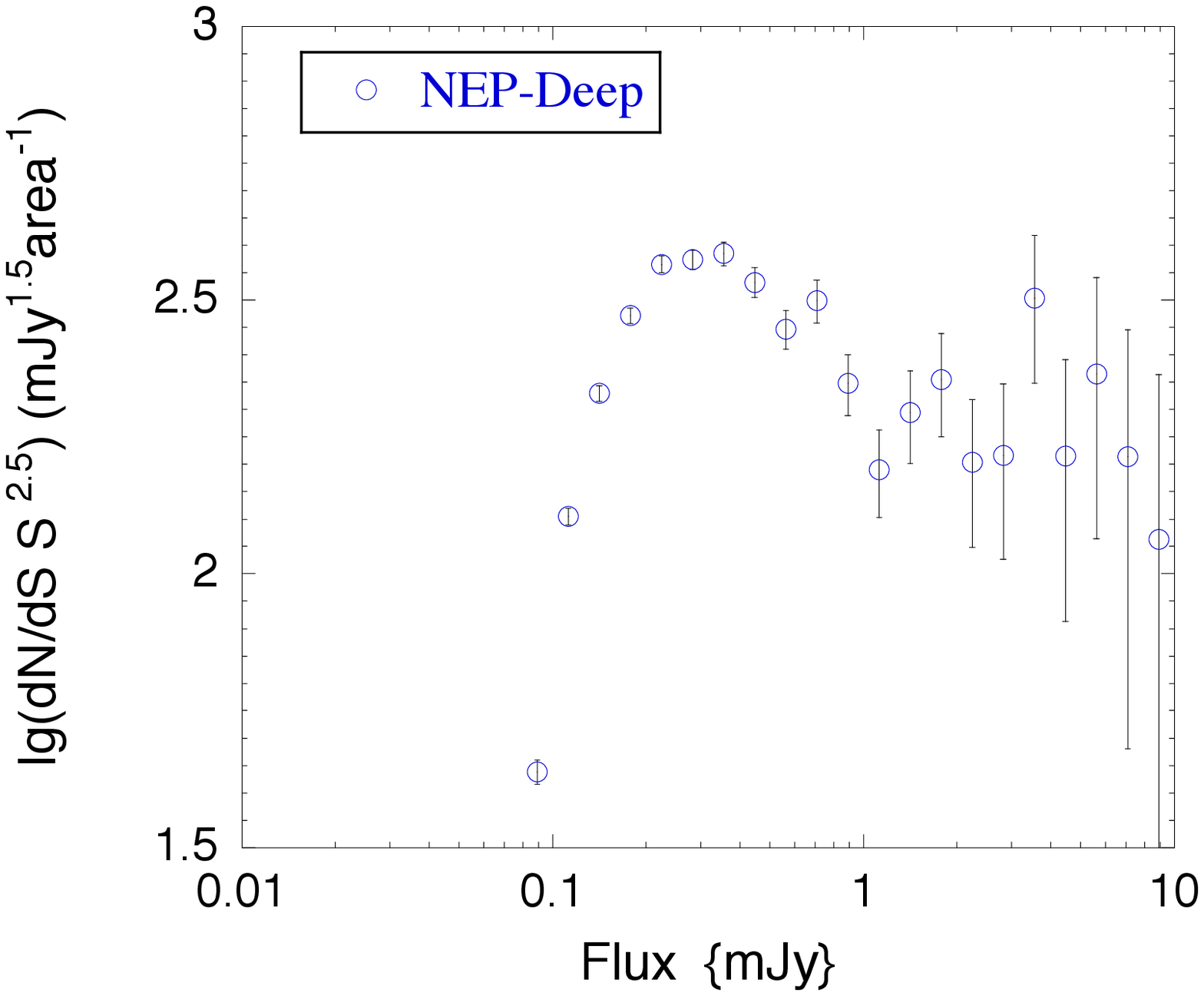,width=7cm}
}
\centerline{
\psfig{ figure= 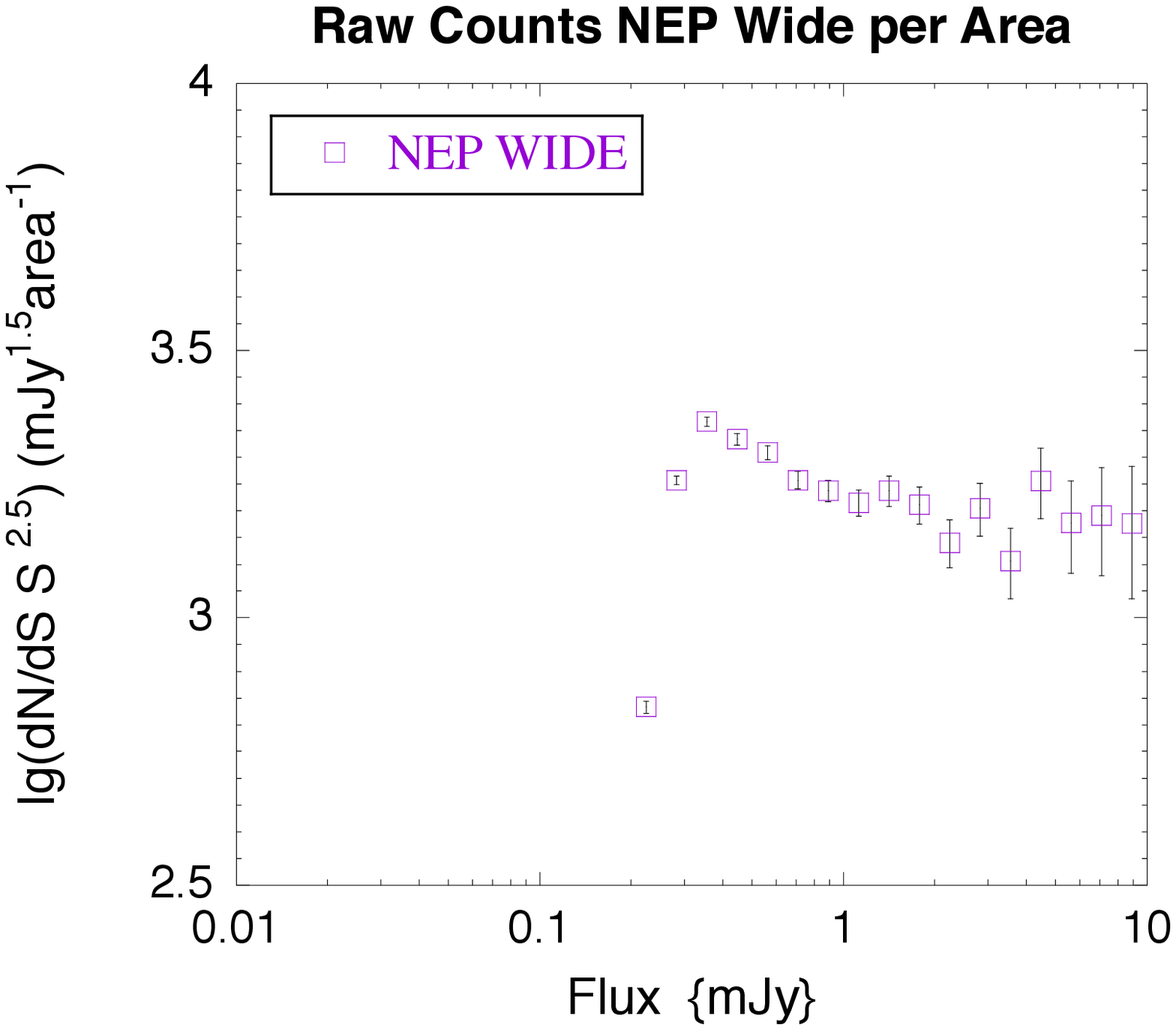,width=7cm}
}
\caption{Raw source counts for the {\it AKARI} NEP-Deep and NEP-Wide surveys in the L18W (18\,$\mu$m) band. {\it Top}: Differential  histogram of the number of sources detected in a flux density bin of size $\Delta\log_{10}S$=0.1 over the areas of the NEP-Deep \& NEP-Wide surveys. {\it Middle}: Raw differential counts for the NEP-Deep survey area in square degrees normalized to a Euclidean Universe.  {\it Bottom}: Raw differential counts for the NEP-Wide survey area in square degrees normalized to a Euclidean Universe.
\label{rawcounts}}
\end{figure}

To calculate the raw source counts from our catalogue we bin the data
in flux density bins of $\Delta\log_{10}S=0.1$. The resulting raw differential
source count histogram is shown in the {\it top panel} of Figure
\ref{rawcounts} calculated over the total areas separately for the NEP-Deep and NEP-Wide surveys respectively. In the NEP-Deep image, we detect a total of
$\sim$5500 sources down to a 5$\sigma$ flux density limit of 80\,$\mu$Jy. From
the NEP-Wide image we detect almost 12,000 sources down to 150$\mu$Jy (5$\sigma$). For the NEP-Deep image, due to nature of its deeper layered
pointing strategy, we find (by visual inspection) we can extract
sources with confidence down to the 3$\sigma$  level, however in the
case of the NEP-Wide image we find that many spurious sources are
selected below the 5$\sigma$ level. The flux density distribution of our
sources peak at 0.16 and 0.3 mJy for the NEP-Deep and Wide surveys
respectively. {\bf The source catalogues are available as additional online material as part of this work. The catalogues are provided separately for the NEP-Deep and NEP-Wide surveys with the first 10 entries in each shown in Tables ~\ref{tab:nepdeepcatalogue} and ~\ref{tab:nepwidecatalogue} respectively.}

A clearer perspective of the source counts is provided by
constructing the Euclidean normalized differential source counts from
the raw source counts.  The Euclidean normalized differential source
counts  given by $(dN/dS) S^{2.5}$ (in units of mJy$^{1.5} $) highlight
any deviation from the flat
(Euclidean) universe expectation.
The normalized differential counts are plotted in
Figure \ref{rawcounts}  for the specific NEP-Deep 
 ({\it middle panel}) and  NEP-Wide  ({\it bottom panel}) areas in square degrees. Errors are Poisson number
errors. We see a characteristic bump starting at flux densities fainter than a milliJansky in the NEP-Deep counts and a peak in the distribution around 200-400\,$\mu$Jy, indicative of evolution in the
source counts. However, the NEP-Wide counts appear much flatter. Note
that in the Figure, due to the different areal coverage, the vertical scales are different  but cover similar orders
of magnitude.

\subsection{Completeness, reliability, stellar fraction and colour corrections}\label{sec:completeness}

Given the circular mosaicing used to create the NEP images, we would expect some uneven coverage, leading to different flux density depths and therefore incomplete samples to a uniform flux density level. Therefore we must make a correction for the  completeness of our survey. 
Moreover an investigation for and correction of any Eddington bias is necessary where, due to noise their is a scatter induced into the flux density of extracted sources relative to their true flux density. Fainter sources suffer larger scatter and because the number of sources at fainter flux densities is larger, the net effect is to effectively boost the number of brighter sources at flux densities around or above the detection limit of the survey.

We estimate the completeness of our source extraction via Monte Carlo simulations by injecting artificial sources into our original images. Our simulations were made as a function of flux density creating simulated sources in flux density bins of $log(S)$=0.1 in individual runs from flux densities of 1000mJy down to 10$\mu$Jy . For each simulation, we injected 200 sources into the original image map and then carried out the same convolution with the matched filter and attempt to recover the sources from the final convolved map. We avoided placing simulated sources next to bright objects in the field. Each simulation at each flux density level consisted of 100 iterations to ensure a statistical robust measure of the completeness of the source extraction.

To evaluate our simulations and derive the completeness correction, following Smail et al. \shortcite{smail95}, Chary et al. \shortcite{chary04}  we create a 2D $P_{ij}$ matrix
with component bins of $S_{in}(i)$ and $S_{out}(i)$. This matrix is populated by the results of our Monte-Carlo simulations for all $i, j$. A  completeness correction is evident since for any given $i$, the normalised sum over all $j$ is $<$1. The matrix was normalised for the sum over $i$ for $j$ to be equal to the number of detected sources in that flux density bin $N_{j}$ . Then the corrected number, $N_{i}$, in each flux density bin $i$ is then given by the sum over $j$ of the re-normalised $P_{ij}$ matrix. 
Compared to a more simplistic approach of comparing number-in with number-out as a function of simulated flux density, the $P_{ij}$ matrix method, has the advantage that it corrects for flux density biases, completeness and scattering due to noise. However, since our measured flux densities are 5$\sigma$ flux densities and we evaluate our counts at the 80$\%$ completeness limit, the method only results in differences of $<$few percent, even at the faintest levels compared to a more simplistic approach.

The completeness correction as a function of flux density, normalized to the survey area is shown in the {\it top panel} of Figure \ref{completeness}. 

The area of a single pixel in our image is 5.645\,arcsec$^2$. Note that although originally the pixels of the IRC-MIR-L array are $2.51\arcsec \times 2.39\arcsec$, resampling carried out during the standard pipeline processing creates square pixels. Due to the masking of noisy regions, image edges etc, the total useful area of the surveys are summed to be 0.623 and 4.967 square degrees for the NEP-Deep and Wide surveys respectively. 

We find that the NEP-Deep survey is around 80$\%$ complete at the $\sim$150\,$\mu$Jy level while the NEP-Wide survey is complete to the same level at around 300\,$\mu$Jy. 

Note that given that the exposure times differ by a factor of 2500/300, we may expect the 80$\%$ completeness sensitivity to differ by a factor of 2.9 rather than a factor of 2. This anomaly may be evidence for the noise dropping off as slower than $\sqrt{t_{exposure}}$ possibly due to instrumental or processing issues or confusion.

In addition to the completeness correction (i.e. whether we include all true sources in our catalogue) we can also investigate the reliability of the catalogues (i.e. whether all sources in the
catalogue are true sources).  We estimate the reliability of our sources at the limit of the survey by carrying out the equivalent source extraction on the negative image. We find  the number of extracted sources from the negative image correspond to a reliability of $\sim$94$\%$ at the limit of our survey (150\,$\mu$Jy) while for the NEP-Wide survey the catalogue appears reliable to the 97$\%$ level at 300\,$\mu$Jy. 
{\bf It should be noted that the negative image is a true test of reliability only if the noise is purely Gaussian. Note that, Figure ~\ref{noise} shows that there is more positive noise than negative noise in the pure noise map, therefore our reliability numbers are likely slightly overestimated.
}
Even in the longer wavelength mid-infrared bands, contamination by stellar sources can affect the bright end of the source counts. To correct for this effect, the stellar fractions as a function of flux density were calculated using the optical data taken by the Canada France Hawaii Telescope (CFHT) and near-infrared data taken by the Kitt Peak National Observatory (KPNO) over the NEP-wide region, which also encompasses the NEP Deep region \cite{hwang07a}. The  {\it AKARI} 18$\mu$m sources were cross matched with CFHT/KPNO data, with stellar sources have stellarity $>$ 0.8 and optical $r'$ band magnitudes $<$ 19 identified during the source extraction.  The stellar sources can also be clearly segregated using the stellarity $>$ 0.8 and optical $r'$ band magnitudes $<$ 19 and the near-infrared colour criteria of H-N2$<$-1.6 in the  H-N2, g-H colour plane (where N2 is the  {\it AKARI} IRC 2.5$\mu$m band). We have also checked our stellar criteria in the H-L18W versus H Color-Magnitude diagram (c.f. Shupe et al. \shortcite{shupe08}, but with somewhat different filter bands). Our criteria agree well (within a few $\%$) and therefore is regarded as sufficient in statistically correcting for stellar sources. In Table \ref{tab:stellarcontribution} the stellar fraction as a function of flux density (bins of lg$(S/mJy)$=0.2) is shown.

Strictly speaking, colour corrections are required for the measured flux densities of galaxies in order to  to account for the different spectra across the IRC L18W band pass filter compared to the stellar calibrators. Unfortunately, no appropriate official colour corrections exist for a power law spectrum of the form $F_{\nu}\propto \nu^{\alpha}$ therefore in order to estimate the colour corrections for our galaxies we simulate a flat spectrum $ \nu F_{\nu}$ = constant and a star sampled on the RayleighÐJeans tail across the L18W 18$\mu$m band pass. We find differences in the flux densities corresponding to a colour correction of  $\sim$3$\%$. We do not apply this colour correction to our source flux densities.

\begin{table}
\caption{ Fraction of stellar contribution to L18W band source counts}
\centering
\begin{tabular}{@{}ll}
\hline
lg(Flux Density) & Stellar fraction    \\
(mJy) &    \\
\hline
1.172&0.25   \\
0.972&0.167   \\
0.772&0.25   \\
0.572&0.214   \\
0.372&0.17   \\
0.172&0.13   \\
-0.028&0.047   \\
-0.228&0.048   \\
-0.428&0.059   \\
-0.628&0.065   \\
-0.828&0.043   \\
-1.028&0.063   \\
-1.228&0.062   \\
\hline
\end{tabular}
\label{tab:stellarcontribution}
\end{table}

\smallskip

\subsection{Final Source Counts}\label{sec:counts}

Applying our completeness correction to the raw source counts in
Figure \ref{rawcounts} we can recover  the final completeness
corrected normalized differential source counts per steradian.
 
\begin{figure}
\centering
\centerline{
\psfig{ figure=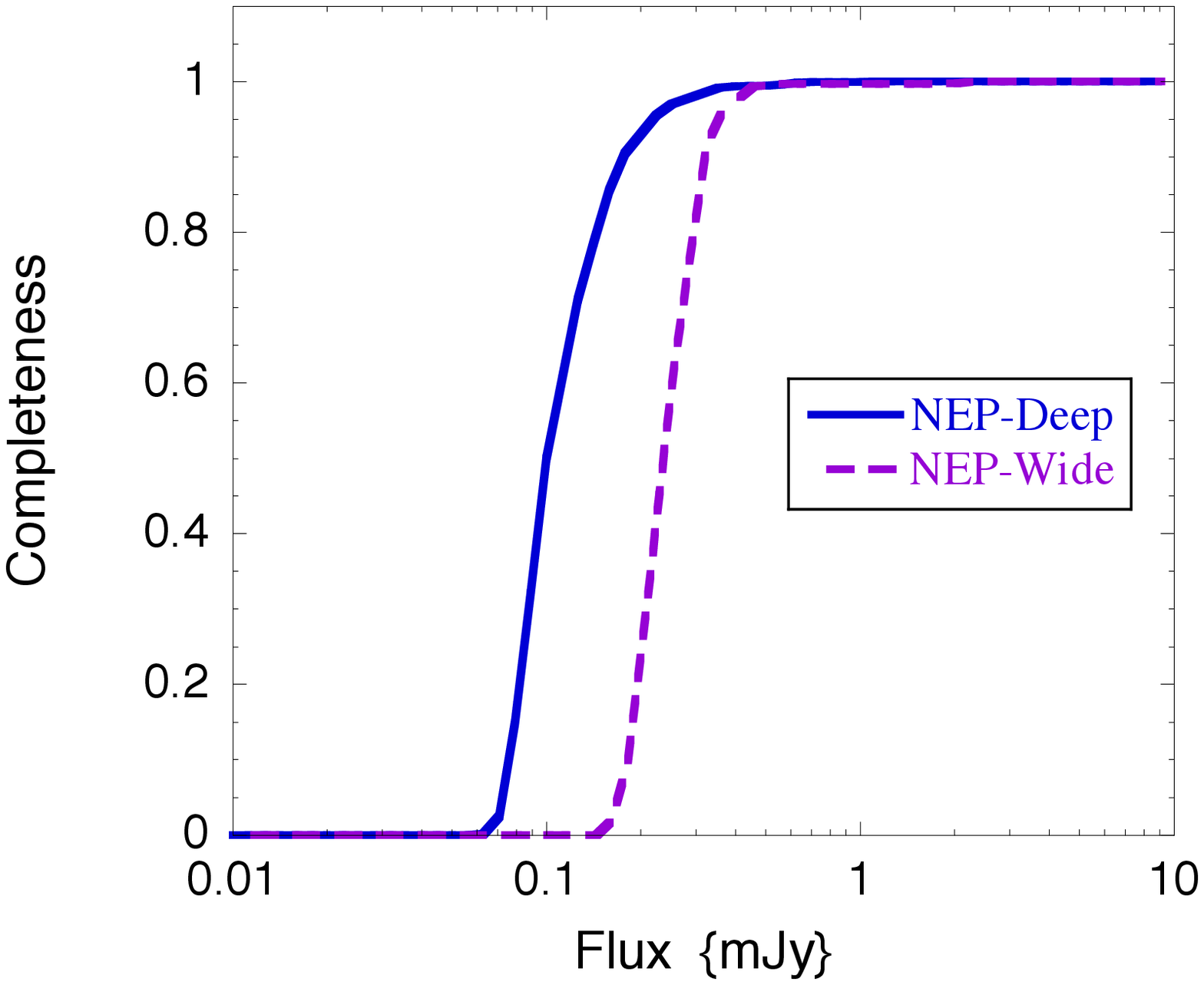,width=7cm}
}
\centerline{
\psfig{ figure=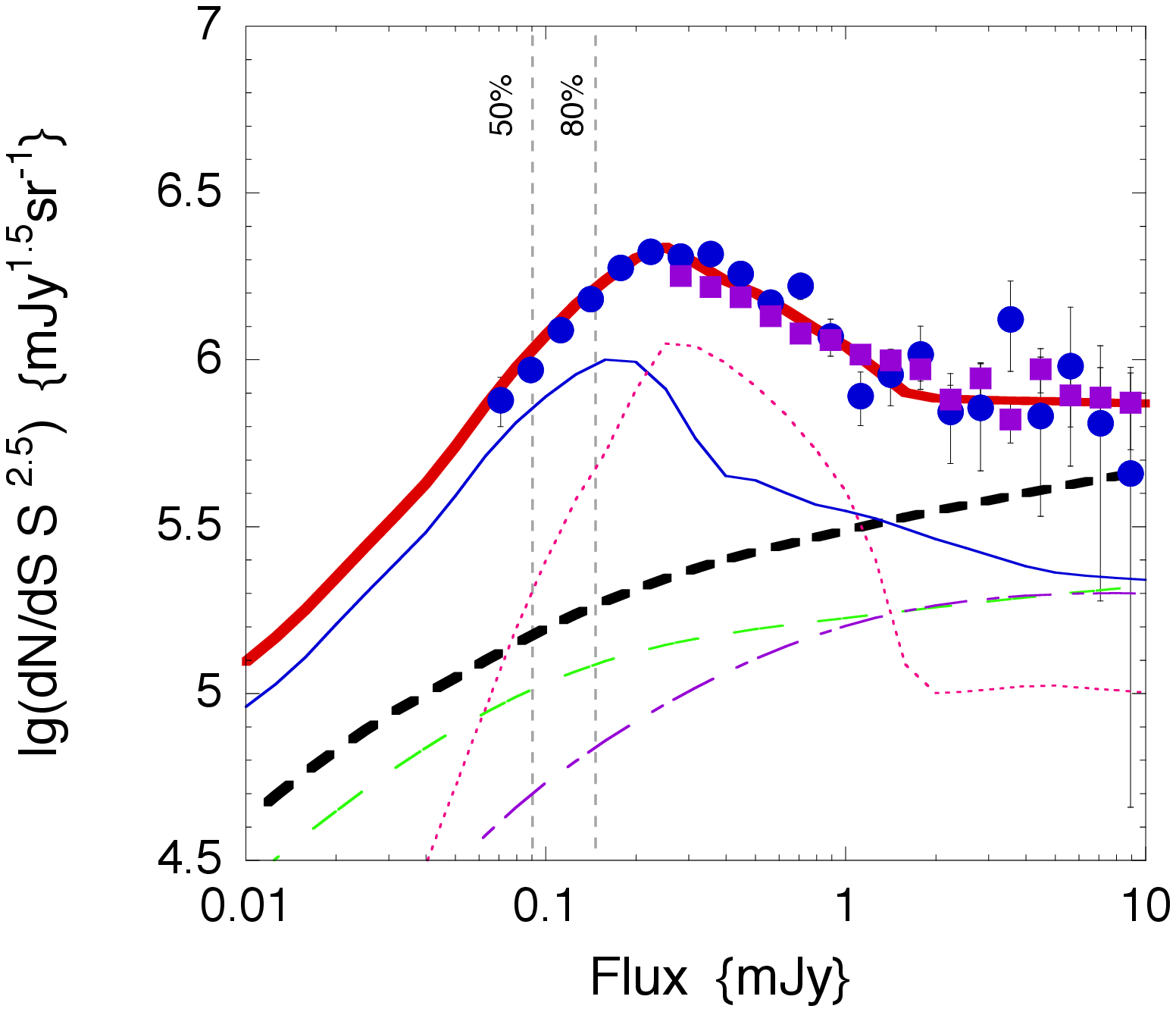,width=7cm}
}
\centerline{
\psfig{ figure=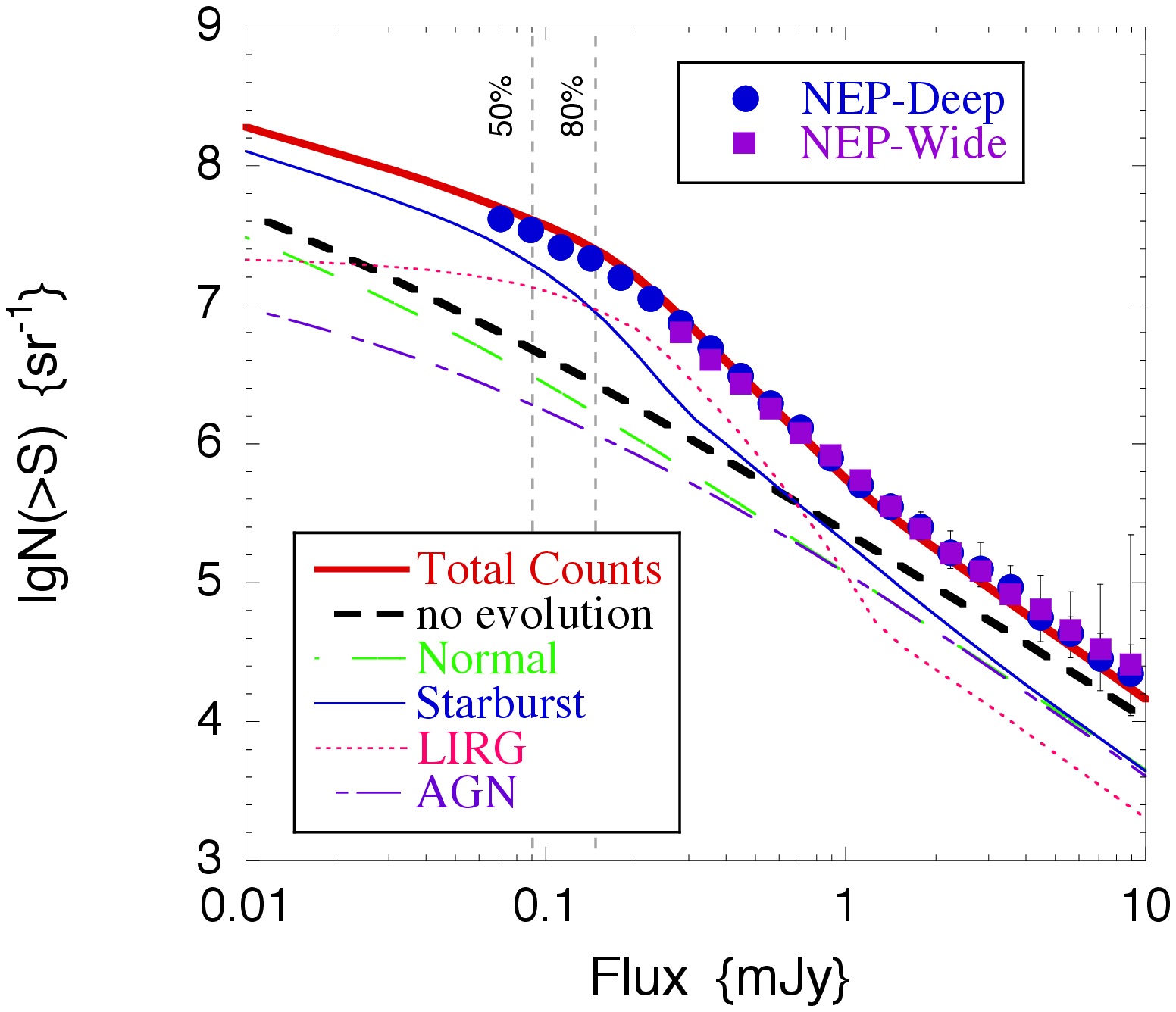,width=7cm}
}
\caption{{\it Top}: Completeness Correction for the  L18W (18\,$\mu$m)
  band {\it AKARI} NEP-Deep and NEP-Wide surveys. {\it Middle}:
  Completeness corrected, stellar subtracted differential source counts per steradian
  normalized for a Euclidean Universe for the {\it AKARI} NEP Deep and
  NEP-Wide surveys. {\it Bottom}: Corresponding integral source counts
  per steradian. The 80$\%$ and 50$\%$ completeness limits for the NEP-Deep survey are shown as {\it vertical-dashed} lines for both the differential and integral counts. For reference we plot the galaxy evolution models of
  Pearson (2005, 2010) as {\it thick-dashed} (no-evolution scenario) and {\it
    thick-solid} (evolving scenario). Model components are shown as {\it thin-dashed} for normal galaxies, {\it thin-solid} for starburst galaxies, {\it thin-dotted} for luminous/ultraluminous galaxies and {\it thin-dot-dash} for AGN. The Total Counts are the sum of all these components. 
\label{completeness}}
\end{figure}

The final counts for the  {\it AKARI} L18W 18\,$\mu$m band are shown in the {\it middle panel} of
Figure \ref{completeness} for the NEP-Deep and NEP-Wide surveys.

 We see the characteristic evolutionary hump below the mJy level
 common to mid-infrared source counts at other wavelengths such as the
 {\it Spitzer} 24\,$\mu$m band (Papovich et
 al. \shortcite{papovich04}, Chary et al. \shortcite{chary04}) and the
 {\it ISO} 15\,$\mu$m band (Aussel et al. \shortcite{aussel99},
 Altieri et al. \shortcite{altieri99}, 
 Serjeant et al. \shortcite{serjeant00}, 
 Gruppioni et
 al. \shortcite{gruppioni02}, Metcalfe et al. \shortcite{metcalfe03},
 Pozzi et al. \shortcite{pozzi04}). The differential source counts in
 the  {\it AKARI} L18W band begin to rise from the Euclidean case
 at a flux density of a mJy with the peak occurring around a flux density
 of 0.25\,mJy. At fainter flux densities ($<0.15$\,mJy) the source
 counts fall away sharply. The counts from the NEP-Deep and NEP-Wide
 surveys are broadly consistent with eachother in the range
 0.3--3\,mJy range where they can be considered to reliably overlap.  

In the   {\it bottom panel} of Figure \ref{completeness} we present
the integral source counts per steradian summed from the differential
source counts, for the NEP-Deep survey and NEP-Wide survey. The source
counts 
between 1\,mJy and 10\,mJy 
exhibit a slope of
$\sim$1.5, i.e. consistent with the non-evolving 
Euclidean universe
steeping to a super-Euclidean slope of $\sim$2 at fainter flux densities
before flattening again to a slope of $\sim$0.8 at flux densities fainter than
0.2\,mJy. At flux densities brighter than $\sim$20mJy the NEP-Wide counts exhibit an
upturn in the counts probably due to both low-number statistics
and an additional uncertain contribution from stellar sources. 
{\bf
 For clarity we also over-plot the  80$\%$ and 50$\%$ completeness limits for the NEP-Deep survey for both the differential and integral counts
( {\it middle panel} and {\it bottom panel} of Figure \ref{completeness}). Note that although the counts are plotted down to flux densities of $\sim$ 70 $\mu$Jy, the completeness at the level is only  $\sim$7$\%$ at this level. 
}

The completeness corrected normalized differential source counts
$\mathrm{d}N/\mathrm{d}S . S^{2.5}$ from Figure  \ref{completeness},
are tabulated in Tables \ref{tab:deepcounts} and  \ref{tab:widecounts}
for the NEP-Deep and NEP-Wide surveys respectively. 

A specific requirement of the {\it AKARI} NEP surveys was to be wide enough to overcome any major contribution from cosmic variance e.g., Somerville et al. \shortcite{somerville04}, contributing as clustering effects on the counts. A thorough analysis of clustering in the NEP field is beyond the scope of this work however  Takeuchi et al. \shortcite{takeuchi01} have shown that the contributed errors on the number counts depend on the angular correlation function and the area of the survey and  provide a formulation for the signal to noise depending on the solid angle of the survey. Drawing a parallel between the NEP survey and  the {\it Spitzer} 24$\mu$m surveys   (Magliocchetti et al. \shortcite{magliocchetti07}, Magliocchetti et al. \shortcite{magliocchetti08}) we estimate errors on the number counts due to clustering of between $<$5$\%$ for redshifts, z$<$1.5.

For reference, in Figure  \ref{completeness} we plot the galaxy evolution model from Pearson \shortcite{cpp05}, \shortcite{cpp10}  for an evolving galaxy scenario (total and component counts) and the no evolution case (the models have not been fitted to the data but are independently derived following Pearson \shortcite{cpp05}). Note that since these are the first source counts presented at 18$\mu$m it is difficult to compare our observations with other easily available contemporary models. Our models are composed of normal quiescent, starburst (L$_{IR}<$10$^{11}$L$\sun$), luminous (LIRG, L$_{IR}>$10$^{11}$L$\sun$) and ultra-luminous (ULIRG, L$_{IR}>$10$^{12}$L$\sun$) galaxy and AGN populations.  The  model assumes a strongly evolving population of luminous infrared galaxies in both luminosity and density and already provides good fits to the 15 and 24\,$\mu$m data from {\it Spitzer}, {\it ISO} \&  {\it AKARI} \cite{cpp10}, including the redshift distribution of the {\it Spitzer} 24$\mu$m population \cite{desai08}.

We see that the differential source counts in the {\it middle panel} of Figure \ref{completeness} rapidly diverge from the no evolution predictions with the observed source counts a factor of three higher than the non-evolving predictions at  $\sim1\,$mJy. The evolving model and the counts agree well at flux densities of 5\,mJy and fainter, with the model reproducing both the upturn in the source counts, the peak between 0.2-0.4\,mJy and the decline to fainter flux densities. At bright flux densities, the model normalization agrees with the source counts out to around 10mJy. At brighter flux densities ($\ge$10\,mJy), errors on the counts are too large to confidently constrain the model. 
Two populations dominate the source counts and together compose the evolutionary hump between 0.2-0.4\,mJy. We see that the upturn in the source counts at around a mJy is caused by the emergence of a population of luminous infrared galaxies (L$_{IR}>$10$^{11}$L$\sun$, e.g. Le Floch et al. \shortcite{lefloch04}) while at fainter flux density levels the counts are composed of less luminous starburst galaxies. AGN only significantly contribute to the source counts at flux densities brighter than 1 mJy (consistent with the fractions observed in the {\it Spitzer} 24$\mu$m population, Brand et al. \shortcite{brand06}).

\smallskip

\section{Discussion and conclusions}\label{sec:conclusions}

We have presented the first galaxy counts at 18\,$\mu$m using the
 {\it AKARI} satellite's surveys at the North Ecliptic Pole, namely
the NEP-Deep ($\sim$0.6 square degrees) and NEP-Wide  ($\sim$5.8
square degrees) surveys. 
We have described our source extraction procedure  using a Wiener
Filtering algorithm to remove background structure and a minimum
variance point source filter method for our source extraction and photometry that delivers
the optimum signal to noise for our extracted sources. We have
constructed robust source catalogues down to 5$\sigma$ flux density levels of
150\,$\mu$Jy and 300\,$\mu$Jy for the NEP-Deep and NEP-Wide surveys
respectively. The calibration of our source flux densities is dependent on the
photometry method and care must be taken if using a different
methodology to the aperture photometry on which the official raw
count-Jansky conversion factors are based. 

A comparison with previous catalogues produced from general mass processing of the NEP-Deep data set (Wada et al.\shortcite{wada08}) concludes that we are achieving significantly  reliable results to fainter flux density levels.  Our 80$\%$ completeness level is a factor of 1.3 deeper compared to Wada et al.\shortcite{wada08}.

The resulting differential and integral source counts from the pair of
surveys in the NEP region are self-consistent over the range where the
flux densities overlap. We reproduce the evolutionary bump in the
source counts between 0.2-0.4\,mJy and the super-Euclidean slopes at
flux densities fainter than a milli-Jansky seen in the previous surveys at 15
and 24\,$\mu$m with {\it ISO} and {\it Spitzer}. The source counts at
flux densities fainter than 1\,mJy are well represented by contemporary source
count models implying that the steep source count slopes are produced
by the rapid evolution of the luminous dusty galaxy population. The {\it AKARI}  L18W
filter  covers the entire wavelength range between the {\it Spitzer} MIPS
24\,$\mu$m and  {\it ISO} ISOCAM 15\,$\mu$m bands (and the 12-22$\mu$m bands of the {\it WISE} surveys: Jarrett et al. \shortcite{jarrett11}, Kennedy \& Wyatt \shortcite{kennedy12}) and may therefore
be probing a correspondingly intermediate population. The high
redshift population revealed by  {\it Spitzer} (Papovich et
al. \shortcite{papovich04}, Le Floch et al. \shortcite{lefloch04})
commonly via detection of the redshifted emission from the
7.7-8.6\,$\mu$m dust features was more or less undetected by  {\it
  ISO} due to the strong K-corrections at higher redshift, however the
broad L18W band should allow detection of both the moderate redshift
{\it ISO} star-forming galaxies and the distant luminous {\it Spitzer}
sources. . Due to the large width $\Delta \lambda$=11.7\,$\mu$m of the passband, for the L18W channel these dust features enter the band at a
redshift of $\sim$0.8 and leave at a redshift of $\sim$2.
More detailed analysis of the full multi-band catalogues from the
{\it AKARI} NEP survey including re-reduction of the images from the raw data \cite{murata13} will be able to link together these populations
and strongly constrain the galaxy evolution models from 2-24\,$\mu$m. 

By integrating the completeness-corrected source counts down to the limit of the NEP-Deep survey at the $\sim$150\,$\mu$Jy level we calculate a lower limit for the
intensity of the 18\,$\mu$m cosmic infrared background of
$\sim$1.4\,nW m$^{-2}$sr$^{-1}$ (1 $\sigma$). Extrapolating to fainter flux densities
using our source count models we predict a total 18\,$\mu$m background
intensity of $\sim$2.6 nW m$^{-2}$sr$^{-1}$ implying that the {\it
  AKARI}  L18W survey at the NEP has resolved approximately 54$\%$ of
the cosmic infrared background at 18\,$\mu$m. Note that these numbers
are consistent with the results of the deep surveys at 24\,$\mu$m
carried out by   {\it Spitzer} which derived a contribution to the
24\,$\mu$m background intensity of 1.9$\pm$0.6\,nW\,m$^{-2}$sr$^{-1}$
with sources fainter than 400\,$\mu$Jy contributing $\sim60\%$  
\cite{papovich04}. The total estimated background intensity at
24\,$\mu$m ($\sim$2.7\,nW\,m$^{-2}$sr$^{-1}$, Dole et al.  ~\shortcite{dole06} ) is also consistent with
the extrapolation to fainter flux densities of our  {\it AKARI}  L18W
observations.

\begin{table}
\caption{ {\it AKARI}  band NEP-Deep survey L18W band  Euclidean normalized differential source counts.}
\centering
\begin{tabular}{@{}lllll}
lg(Flux Density) & Counts& \multicolumn{2}{c}{Errors} & Completeness  \\
&  $\mathrm{d}N/\mathrm{d}S . S^{2.5}$ &(low) & (high) & \\
 (mJy) & (mJy$^{1.5}$sr$^{-1}$) & \multicolumn{2}{c}{(mJy$^{1.5}$sr$^{-1}$)} &  \\
\hline
-1.15	&	5.88	&	5.80	&	5.97	& 0.07 \\
-1.05	&	5.97	&	5.95	&	5.99	& 0.33 \\
-0.95	&	6.09	&	6.11	&	6.13	& 0.51 \\
-0.85	&	6.18	&	6.13	&	6.19	& 0.77 \\
-0.75	&	6.27	&	6.26	&	6.29	& 0.89 \\
-0.65	&	6.32	&	6.30	&	6.33	& 0.96 \\
-0.55	&	6.30	&	6.29	&	6.32	& 0.98 \\
-0.45	&	6.32	&	6.29	&	6.33	& 0.99 \\
-0.35	&	6.26	&	6.22	&	6.28	& 1.00\\
-0.25	&	6.17	&	6.13	&	6.20	& 1.00\\
-0.15	&	6.22	&	6.18	&	6.26	& 1.00\\
-0.05	&	6.07	&	6.01	&	6.12	& 1.00\\
0.05	&	5.89	&	5.80	&	5.96	& 1.00\\
0.15	&	5.96	&	5.86	&	6.03	&1.00 \\
0.25	&	6.01	&	5.91	&	6.10	& 1.00\\
0.35	&	5.84	&	5.68	&	5.96	& 1.00\\
0.45	&	5.86	&	5.67	&	5.98	& 1.00\\
0.55	&	6.12	&	5.96	&	6.23	& 1.00\\
0.65	&	5.83	&	5.53	&	6.00	& 1.00\\
0.75	&	5.98	&	5.68	&	6.15	& 1.00\\
0.85	&	5.81	&	5.27	&	6.04	& 1.00\\
0.95	&	5.66	&	4.66	&	5.96	& 1.00\\
\hline
\end{tabular}
\label{tab:deepcounts}
\end{table}

\begin{table}
\caption{ {\it AKARI}  band NEP-Wide survey L18W band Euclidean normalized differential source counts.}
\centering
\begin{tabular}{@{}lllll}
lg(Flux Density) & Counts& \multicolumn{2}{c}{Errors}   & Completeness \\
&  $\mathrm{d}N/\mathrm{d}S . S^{2.5}$ &(low) & (high)& \\
 (mJy) & (mJy$^{1.5}$sr$^{-1}$) & \multicolumn{2}{c}{(mJy$^{1.5}$sr$^{-1}$)} & \\
\hline
-0.55	&	6.25	&	6.24	&	6.26	& 0.77 \\
-0.45	&	6.22	&	6.21	&	6.23	& 0.95 \\
-0.35	&	6.19	&	6.18	&	6.20	& 1.00\\
-0.25	&	6.13	&	6.12	&	6.14	& 1.00\\
-0.15	&	6.08	&	6.06	&	6.09	& 1.00\\
-0.05	&	6.06	&	6.03	&	6.07	& 1.00\\
0.05	&	6.02	&	5.99	&	6.04	& 1.00 \\
0.15	&	5.99	&	5.97	&	6.02	& 1.00 \\
0.25	&	5.97	&	5.93	&	6.00	& 1.00 \\
0.35	&	5.88	&	5.83	&	5.92	& 1.00 \\
0.45	&	5.94	&	5.89	&	5.99	& 1.00 \\
0.55	&	5.82	&	5.75	&	5.88	& 1.00 \\
0.65	&	5.97	&	5.90	&	6.03	& 1.00 \\
0.75	&	5.89	&	5.79	&	5.97	& 1.00 \\
0.85	&	5.88	&	5.77	&	5.97	& 1.00 \\
0.95	&	5.87	&	5.73	&	5.97	& 1.00 \\
1.05	&	5.95	&	5.78	&	6.07	& 1.00 \\
\hline
\end{tabular}
\label{tab:widecounts}
\end{table}

\smallskip

\section{Acknowledgements}
Chris Pearson acknowledges his Bridge Fellowship from the Japan Society for the Promotion of Science, under which some of this work was carried out. MI acknowledges the support from the National Research Foundation of Korea (NRF) grant, No. 2008-0060544 funded by the Korea government (MSIP). SS, CP (The Open University) acknowledge support from STFC grants ST/J001597/1 and ST/G002533/1 and the Royal Society (2006/R4-IJP). HML and SJK were  supported by the NRF grant No. 2012R1A4A1028713 funded by the Korea Government (MSIP).
The {\it AKARI} Project is an infrared mission of the Japan Space Exploration Agency (JAXA) Institute of Space and Astronautical Science (ISAS), and is carried out with the participation of mainly the following institutes; Nagoya University, The University of Tokyo, National Astronomical Observatory Japan, The European Space Agency (ESA), Imperial College London, University of Sussex, The Open University (UK), University of Groningen / SRON (The Netherlands), Seoul National University (Korea). The far-infrared detectors were developed under collaboration with The National Institute of Information and Communications Technology. This research made use of Tiny Tim/Spitzer, developed by John Krist for the Spitzer Science Center. The Center is managed by the California Institute of Technology under a contract with NASA. The authors would like to thank the anonymous referee for their comments and recommendations that have improved this work.



\appendix

\section[]{Source Catalogues}

The source catalogues are available as additional online material as part of this work. The catalogues are provided separately for the NEP-Deep and NEP-Wide surveys with the first 10 entries in each shown in Tables ~\ref{tab:nepdeepcatalogue} and ~\ref{tab:nepwidecatalogue} respectively.

\begin{table*}
\caption{ Source catalogue for NEP-Deep survey. Table columns are Right Ascension, Declination (J2000, decimal degrees), flux density and flux density error (mJy). The complete catalogue is available online.}
\centering
\begin{tabular}{@{}llll}
R.A. & Dec.& Flux Density & Flux Density Error  \\
 (deg.) & (deg.) & (mJy) &  (mJy) \\
\hline
269.0355448	&	66.08776218	&	0.21411	&	0.06870	\\
269.0648752	&	66.08969796	&	0.21116	&	0.05688	\\
269.0779578	&	66.09495669	&	0.20007	&	0.05785	\\
269.0193719	&	66.10164498	&	0.26344	&	0.05695	\\
269.0324457	&	66.10624754	&	0.15348	&	0.05693	\\
269.1009026	&	66.10745694	&	0.19675	&	0.05904	\\
269.0471359	&	66.1088665	&	0.50443	&	0.05686	\\
269.0275789	&	66.10889441	&	0.28341	&	0.05752	\\
268.9086087	&	66.10967099	&	0.29854	&	0.05705	\\
269.0618293	&	66.11148406	&	0.16113	&	0.05682	\\
\hline
\end{tabular}
\label{tab:nepdeepcatalogue}
\end{table*}

\begin{table*}
\caption{ Source catalogue for NEP-Wide survey. Table columns are Right Ascension, Declination (J2000, decimal degrees), flux density and flux density error (mJy). The complete catalogue is available online.}
\centering
\begin{tabular}{@{}llll}
R.A. & Dec.& Flux Density & Flux Density Error  \\
 (deg.) & (deg.) & (mJy) &  (mJy) \\
\hline
270.074791	&	65.20182091	&	0.97093	&	0.07878	\\
270.0543152	&	65.21106171	&	0.62312	&	0.07860	\\
269.9818059	&	65.22753773	&	0.35827	&	0.07860	\\
269.6461222	&	65.22760952	&	0.45081	&	0.07623	\\
269.5625606	&	65.22932738	&	0.35855	&	0.07653	\\
270.1457166	&	65.23151435	&	0.47485	&	0.07910	\\
269.8635498	&	65.23467804	&	0.27454	&	0.05387	\\
270.6769667	&	65.23429968	&	0.39382	&	0.08360	\\
270.0732105	&	65.23549074	&	0.30702	&	0.05577	\\
268.9540795	&	65.23129776	&	0.54024	&	0.07575	\\\hline
\end{tabular}
\label{tab:nepwidecatalogue}
\end{table*}

\bsp 

\label{lastpage}

\end{document}